\documentclass{statsoc}

\usepackage[a4paper]{geometry}
\usepackage{graphicx}
\usepackage[textwidth=9em,textsize=small]{todonotes}

\usepackage{times}
\usepackage{epsfig}
\usepackage{natbib}
\usepackage{graphicx,color}
\usepackage{amsmath}
\usepackage{amssymb}
\usepackage{multirow}

\def\argmin{\mathop{\rm argmin}}

\newcommand{\s}{\ensuremath{\mathbb{S}}}

\newcommand{\real}{\ensuremath{\mathbb{R}}}

\newcommand{\ltwo}{\ensuremath{\mathbb{L}^2}}

\def\sl{\mathop{\mathfrak s\mathfrak l}\nolimits}

\newtheorem{algorithm}{Algorithm}
\newtheorem{thm}{Theorem}

\def\argmin{\mathop{\rm argmin}}

\newcommand{\cC}{\ensuremath{\mathbb{C}}}

\newtheorem{defn}{Definition}
\newtheorem{lemma}{Lemma}

\newtheorem{prop}{Proposition}

\title{Phase-Amplitude Separation and Modeling of Spherical Trajectories}

\author{Zhengwu Zhang}
\address{Statistical and Applied Mathematical Sciences Institute, Durham, USA}
\author{Eric Klassen}
\address{Department of Mathematics, Florida State University, Tallahassee, USA}
\author[Z. Zhang, E. Klassen, A. Srivastava]{Anuj Srivastava}
\address{Department of Statistics, Florida State University, Tallahassee, USA}

\begin{document}

\begin{abstract}
This paper studies the problem of separating phase-amplitude components in sample paths of a spherical 
process (longitudinal data
on a unit two-sphere). Such separation is essential for efficient modeling and 
statistical analysis of spherical longitudinal data in a manner that is invariant to any phase variability. 
The key idea is to represent each path or trajectory with a pair of variables, a starting point and a 
Transported Square-Root Velocity Curve (TSRVC). A TSRVC is a curve in the tangent (vector) space at the starting point
and has some important invariance properties under the $\ltwo$ norm. 
The space of all such curves forms a vector bundle and 
the $\ltwo$ norm, along with the standard Riemannian metric on $\s^2$, provides a natural metric on 
this vector bundle. 
This invariant representation allows for separating phase and amplitude components in given data, using a template-based
idea. Furthermore, 
the metric property is useful in deriving computational  procedures for clustering, mean computation, 
principal component analysis (PCA), and modeling. 
This comprehensive framework is demonstrated using two datasets: a set of bird-migration trajectories and 
a set of hurricane paths in the Atlantic ocean. 
\end{abstract}

\section{Introduction}
\begin{figure}C
\begin{center}
\begin{tabular}{cc}
\includegraphics[height=1.5in]{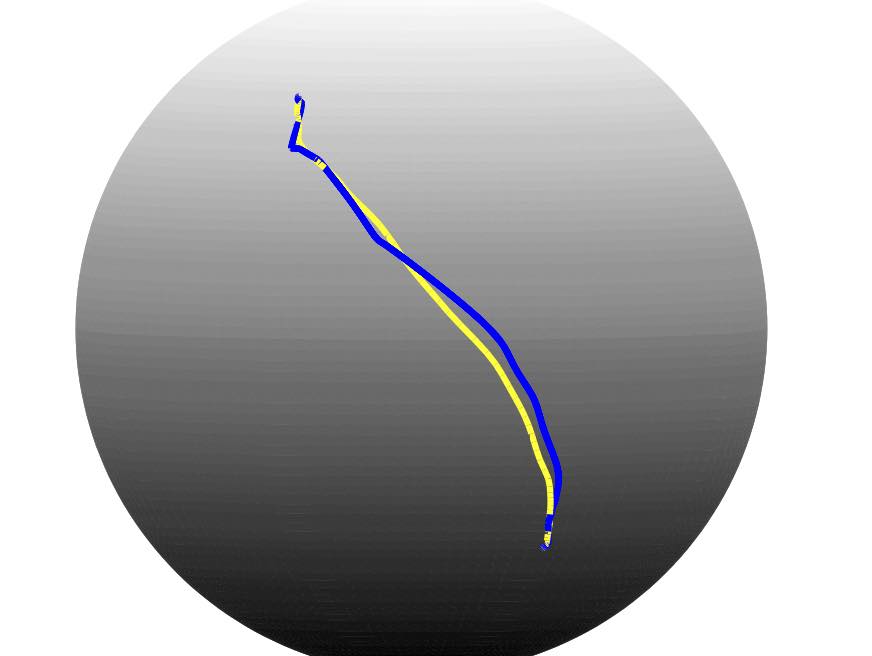}&
\includegraphics[height=1.5in]{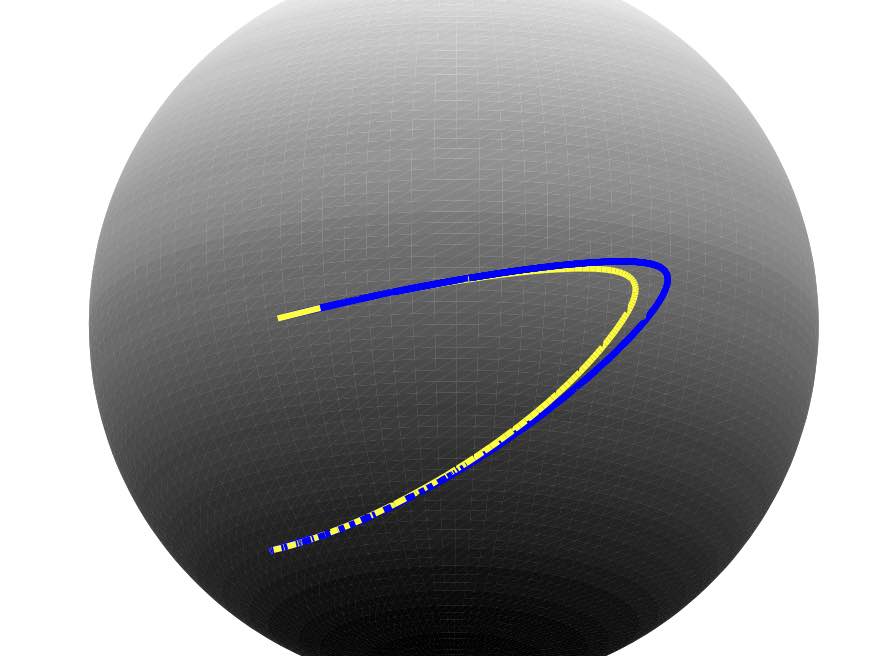}\\
\end{tabular}
\caption{Cross-sectional mean of simulated trajectories. In each plot, blue lines show $10$ trajectories with only phase differences, and the yellow line shows the cross-sectional mean of the 10 trajectories.}\label{fig:crossmean} 
\end{center}
\end{figure}

Many dynamical systems can be characterized as temporal evolutions of a state variable over a nonlinear manifold $\cal M$. 
Given discrete observations, or sample paths,  of such systems,
the goal is to perform statistical modeling, prediction and parameter estimation. 
For instance, one may be interested in defining and computing statistical summaries, 
i.e. mean and covariance, of the given sample paths. 
Also, one can use these estimated summaries in discovering dominant modes of variability and 
performing dimension reduction, e.g. using
PCA.  Another application is to cluster and classify 
trajectories into some pre-determined classes. 
These problems are complicated for several reasons. One is, of course, the nonlinear geometry of ${\cal M}$, 
which may not allow for standard multivariate statistics to be applied directly. 
Secondly,  very often the data is collected in presence of phase variability, which further complicates
data analysis. Roughly speaking, the phase variability corresponds to a lack of registration 
of time points along trajectories.  If two trajectories follow the same sequence of points on ${\cal M}$ but at different times, then they are 
said to have different phases but the same amplitude.
If these phase variabilities are not taken into account, they lead to 
loss of structure in the mean calculation, artificial inflation of variance, and introduction of spurious principal components
\citep{marron-etal:StatSci:2015}. 
In Fig. \ref{fig:crossmean} we illustrate this effect, where the yellow line is the cross-sectional mean of $10$ trajectories on $\s^2$ plotted in blue lines. The overlapping blue trajectories follow the same sequence of points but have different time parameterizations, and 
the mean ends up passing through a different sequence of points.   
Any resulting statistical model can be rendered ineffective due to this problem. 
In order to overcome this issue, one has to register the trajectories
or, in other words, perform phase-amplitude separation.  

Phase-amplitude separation for Euclidean data is well studied now. 
See, for example, the papers  \citep{muller-JASA:2004,kneip-ramsay:2008,muller-biometrika:2008,srivastava-etal-function:2011,Tucker:2013,marron-etal:StatSci:2015,marron-etal-EJS:2014} for scalar functions and  \citep{younes-distance,younes-michor-mumford-shah:08,anuj2011}
for curves in $\real^2$ and higher dimensions.
There is limited discussion when the longitudinal data lies on nonlinear domains, see
\citep{Kume:2007,su2014,Zhang2015,brigant-arXiv:2015,brigant-arXiv-2016} for some general frameworks.
However, if the domain is a canonical one, such as
${\cal M }= \s^2$, it will be very useful to particularize these general solutions into efficient procedures
using the geometry of $\s^2$. 
For spherical longitudinal data, one expects a more 
efficient solution if the detailed geometry of $\s^2$ is incorporated in the solution. The unit sphere plays an 
important role in statistical analysis of directional data \citep{mardia2008} and geophysical phenomena \citep{kendall-arxiv:2014}.
To our knowledge, there is no current paper that advances the solution for phase-amplitude 
separation in spherical trajectories explicitly.  

\begin{figure}[h!]
\begin{center}
\begin{tabular}{cc}
\includegraphics[height=1.7in]{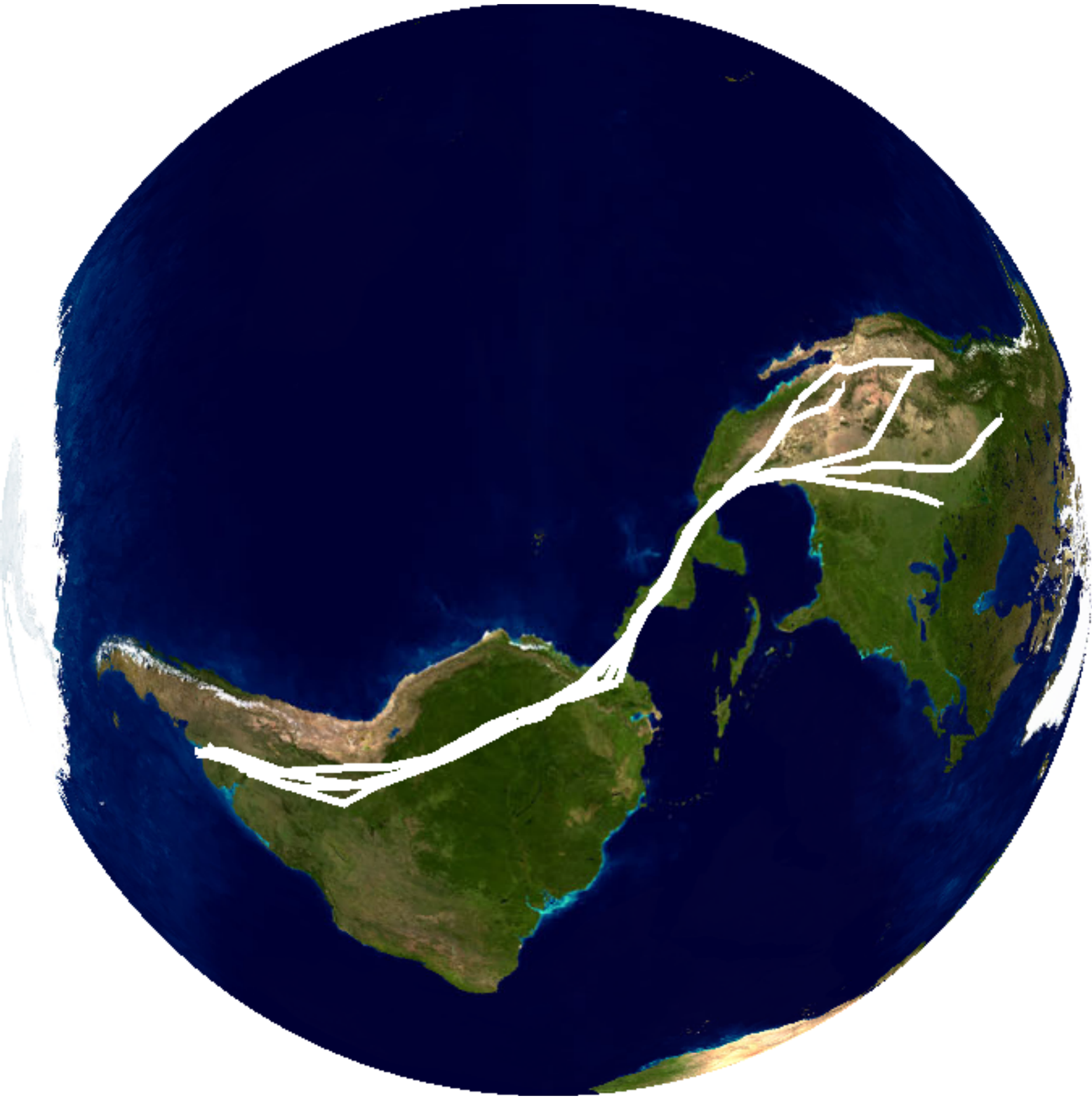}&
\includegraphics[height=1.7in]{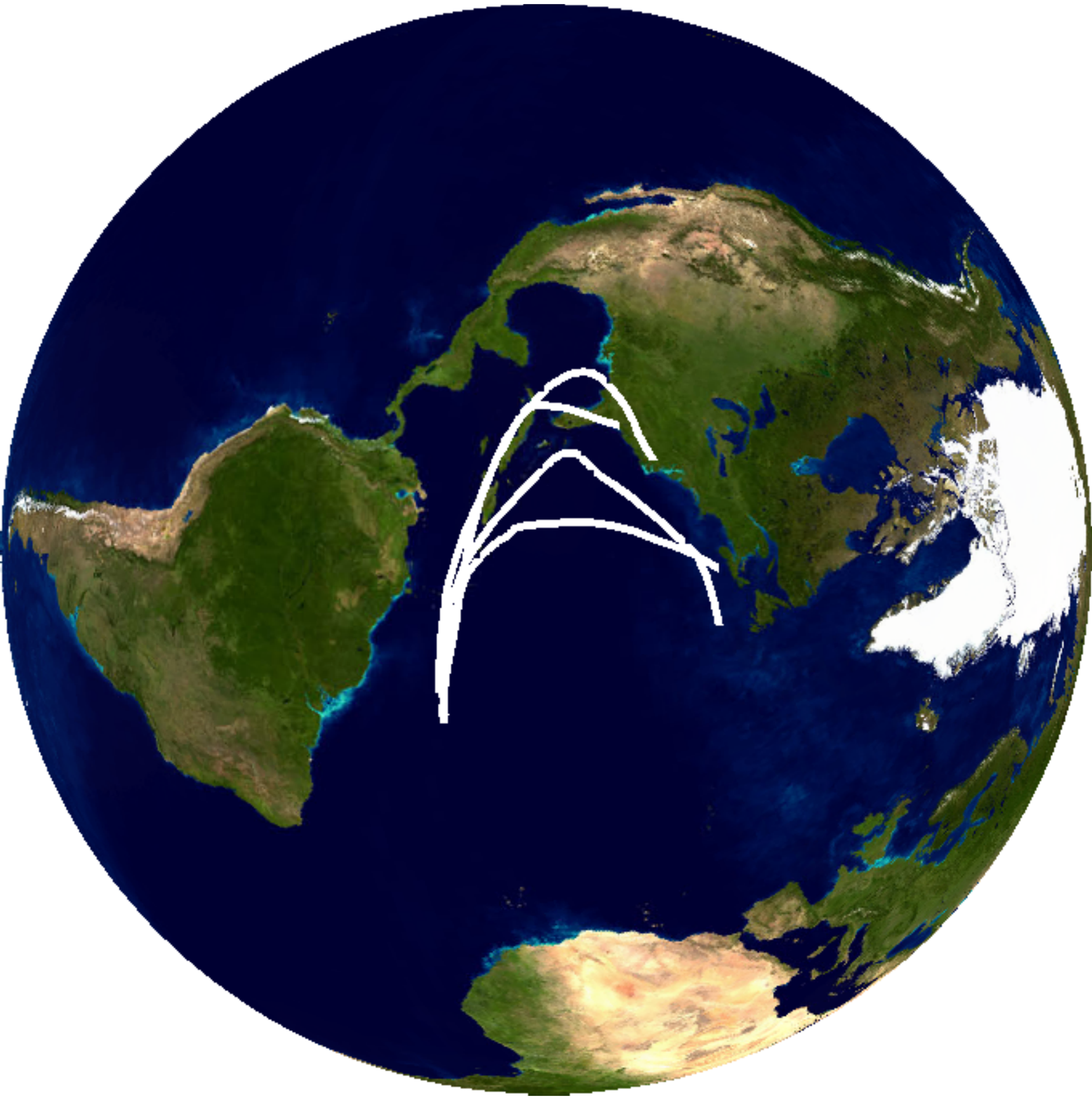}\\
Swainson's hawk migration paths & Hurricane tracks\\
\end{tabular}
\caption{Bird migration paths and hurricane tracks as trajectories on $\s^2$.} \label{application_mot}
\end{center}
\end{figure}

To motivate this problem, we consider two datasets shown in Fig. \ref{application_mot}. The left side shows migration
paths of a type of bird called {\it Swainson hawk} and the right side shows some tracks for hurricanes originating in 
the Atlantic ocean.  Since these tracks are expressed in geographical coordinates, it is natural
to treat the underlying domain as  $\s^2$. During migration,  flocks of birds exhibit 
tremendous variability in travel rates over long distances -- they take very similar routes but their speed
patterns along those routes can vastly differ. 
Similarly, different hurricanes evolve at different temporal 
rates even if they go through similar geographical coordinates. If we compute their cross-sectional statistics, i.e. 
point-wise mean and covariance with the given time labels, we notice that the means are not representative of 
individual trajectories and the variances are artificially large despite the trajectories being very similar.  

To make these concepts precise, we develop some notation first. Let 
$\alpha: [0,1] \to \s^2$ be a trajectory, perhaps generated as 
observations of a dynamical system on the time interval $[0,1]$. Also, let 
 $\gamma:[0,1] \to [0,1]$ be a positive diffeomorphism such that $\gamma(0) = 0$ and $\gamma(1) = 1$. Such 
 a $\gamma$ 
 plays the role of a time-warping function, or a phase function, so that the composition $\alpha \circ \gamma$
 is now a time-warped or re-parameterized version of $\alpha$. In other words, the trajectory $\alpha \circ \gamma$ 
 has the same amplitude as that of $\alpha$, but a different phase.  With this notation, we can specify the following
 registration problems. 
 \begin{enumerate}
 \item {\bf Pairwise Registration}: 
 For any two trajectories $\alpha_1, \alpha_2: [0,1] \to \s^2$, the process of registration of
$\alpha_1$ and $\alpha_2$ is to find a time warping $\gamma$ such that $\alpha_1(t)$ is 
optimally registered to $\alpha_2(\gamma(t))$ for all $t \in [0,1]$. In order to ascribe a meaning to optimality, we
have to develop a well-defined criterion. 

\item {\bf Multiple Registration or Phase-Amplitude Separation}: This problem 
can naturally be extended to more than two trajectories: let $\alpha_1, \alpha_2,\dots, \alpha_n$ be $n$ trajectories on $\s^2$, and we
want to find out time warpings $\gamma_1, \gamma_2, \dots, \gamma_n$ such that for all $t$, the variables $\{\alpha_i(\gamma_i(t))\}_{i=1}^n$ 
are optimally registered. 
The function $\alpha_i(\gamma_i(t))$ represents the {\it amplitude} and $\gamma_i$ is called the {\it phase} of 
$\alpha_i$. If we have a solution from pairwise registration, it can be extended to the multiple alignment problem as 
follows -- for the given trajectories, first define a {\it template} trajectory and then align each given trajectory to this 
template in a pairwise fashion.  One way of defining this template is to use the mean of given trajectories 
under an appropriately chosen metric. 
\end{enumerate}

\subsection{Past Work \& Their Limitations}

Let $d_m$ denote the geodesic distance resulting from the chosen Riemannian metric on $\s^2$. It can be shown that
the quantity $\int_0^1 d_m(\alpha_1(t), \alpha_2(t)) dt$ forms a proper distance on the set $\{\alpha:[0,1] \rightarrow \s^2 \}$, the space of 
all trajectories on $\s^2$. For example, \citet{kendall-arxiv:2014} uses this metric, combined with the arc-length distance on $\s^2$, to cluster hurricane data. However, this metric is not immune to different temporal evolutions of hurricane tracks. Handling this variability requires temporal alignment before or during comparison. 
It is tempting to use the following modification of this distance to align two trajectories: 
$$
\inf_{\gamma \in \Gamma} \left( \int_0^1 d_m(\alpha_1(t), \alpha_2 (\gamma(t))) dt \right)\ ,
$$
but this can lead to degenerate solutions (also known as the {\it pinching problem}, described for real-valued functions in 
\citep{marron-etal:StatSci:2015}). While the degeneracy can be avoided
using a regularization penalty on $\gamma$, some of the other problems remain, including the fact that the solution is not symmetric. 

A recent paper \citet{su2014} developed the concept of elastic trajectories to deal with the 
phase variability in manifold-valued trajectories. 
Here, a trajectory on $\s^2$ is represented by its transported square-root vector field (TSRVF) defined as: 
$h_{\alpha}(t) =  \left( \dot{\alpha}(t) \over \sqrt{ | \dot{\alpha}(t) |} \right)_{\alpha(t) \rightarrow c}  \in T_{c}(\s^2)$,
where $c$ is a pre-determined reference point on $\s^2$ and $\rightarrow$ denotes a parallel transport of the 
vector $\dot{\alpha}(t)/\sqrt{ | \dot{\alpha}(t) |}$ from the point $\alpha(t)$ to $c$ along a geodesic path. This way a trajectory can be mapped
into the tangent space $T_c(\s^2)$ and one can compare/align them using the $\ltwo$ norm on that vector space. 
More precisely, for any two spherical trajectories $\alpha_1$ and $\alpha_2$,  
the quantity $\inf_{\gamma} \| h_{\alpha_1} - h_{\alpha_2 \circ \gamma}\|$ provides not only a 
criterion estimating the phase ( $\gamma$) but it is also a proper metric for averaging and other statistical analyses.  
The main limitation of this framework is that the choice of reference point, $c$, is left arbitrary. It is possible that the 
results can change with $c$ and make the analysis difficult to interpret. A related, and bigger issue, is that the
transport of tangent vectors $\dot{\alpha}(t)/\sqrt{ | \dot{\alpha}(t) |}$  to $c$ can introduce large distortion, especially when the trajectories
are far from $c$ on the manifold $\s^2$. Consider an example where $c$ is the north pole and two
points lying on the great circle passing through the two poles 
but on the opposite sides and close to the south pole. If we take two 
tangent vectors at these two points, that are similar in direction and magnitude, and transport them 
individually to $c$ as discussed above, the resulting vectors at $c$ will be in opposite directions.

\subsection{Our Approach}

In this paper, we overcome the problems associated with TSRVF of \citet{su2014} using a more intrinsic approach.
Here
the trajectories are represented by curves that will not be transported to a global reference point. 
For a trajectory $\alpha$, the reference
point is chosen to be its starting point $\alpha(0) \in \s^2$, and the transport is performed along the trajectory itself. In other words, for each 
$t$, the square-root velocity vector $\dot{\alpha}(t)/\sqrt{| \dot{\alpha}(t) |}$ is transported along $\alpha$ to the tangent space
$T_{\alpha(0)}(\s^2)$. This results in a 
curve in the tangent space $T_{\alpha(0)}(\s^2)$ and our goal is to compare and analyze such curves. 
However, for different trajectories, the starting points are different and we need a proper metric to be 
able to compare these curves in different tangent spaces. We define a natural metric on the representation space of 
such curves, and use it 
for comparing, averaging, and modeling such curves. Similar to the earlier work, this framework is invariant to the re-parameterization 
of trajectories, and provides a natural solution for performing phase-amplitude separation.

The rest of the paper is organized as follows. In Appendix A, 
we introduce a basic Riemannian structure on $\s^2$ to facilitate our analysis of trajectories on $\s^2$. 
In Section 2, we lay out our framework of analyzing spherical trajectories, including a computational 
solution for their phase-amplitude separation. Some statistical methods of modeling trajectories on $\s^2$ are presented in Section 3. 
Section 4 presents the experimental results on both simulated and real data, and the paper ends with a brief discussion 
in Section 5.

\section{Analysis of Trajectories on $\s^2$}
We have summarized briefly the Riemannian structure and certain 
geometric quantities on $\s^2$ in the appendix.  Now we focus on the problem of analyzing trajectories on $\s^2$.

\subsection{Mathematical Representation of Trajectories on $\s^2$}
Let $\alpha$ denote a smooth trajectory on $\s^2$ and ${\cal F}$ denote the set of all such trajectories: ${\cal F} =
\{ \alpha:[0,1] \to \s^2 | \alpha\ \ \mbox{is smooth}\}$. Define
$\Gamma$ to be the set of all orientation preserving diffeomorphisms
of $[0,1]$: $\Gamma=\{\gamma:[0,1]\to [0,1]|\gamma(0)=0,\
\gamma(1)=1,\ \gamma\,$ is a diffeomorphism$\}$. It is important to
note that $\Gamma$ forms a group under the composition operation. If
$\alpha$ is a trajectory on $\s^2$, then $\alpha \circ \gamma$ is a
trajectory that follows the same sequence of points as $\alpha$ but
at the evolution rate governed by $\gamma$. More technically, the
group $\Gamma$ acts on ${\cal F}$, ${\cal F} \times \Gamma \to {\cal
F}$, according to  $(\alpha*\gamma)=\alpha\circ\gamma$.

We now introduce a new representation of trajectories that forms the foundation of our statistical analysis. 
Given a trajectory $\alpha$, let $(v)_{\alpha(t_1) \rightarrow \alpha(t_2)}$ denote the parallel transport of any vector
$v \in T_{\alpha(t_1)}(\s^2)$ along $\alpha$ from $T_{\alpha(t_1)}(\s^2)$ to $T_{\alpha(t_2)}(\s^2)$.

\begin{defn}
For a smooth spherical trajectory $\alpha: [0, 1] \rightarrow \s^2$, with its starting point $\alpha(0)=p$ 
and velocity vector field $\dot{\alpha}(t)$, define its transported square-root vector curve (TSRVC) to
be a scaled parallel transport of  $\dot{\alpha}(t)$ along $\alpha$ to the starting point $p$ according to: $
q(t) = \left( { \dot{\alpha}(t)  \over \sqrt{ | \dot{\alpha}(t) |} } \right)_{\alpha(t) \rightarrow p} \in T_{p}(\s^2) \ $, where
$| \cdot |$ denotes the $\ltwo$ norm defined through the Riemannian metric on $\s^2$. 
\end{defn}
Note that this representation is different from the one in \citet{su2014} in two aspects: (1) The reference point is chosen as the starting point of the trajectory.  (2) The parallel transport of the square-root velocity vector is along the trajectory $\alpha$ itself to the starting point, 
similar to the idea discussed in \citep{Kume:2007}. These two changes reduce the distortion in representation relative to the parallel transport of \citet{su2014} to a far away reference point. 

The TSRVC representation maps a trajectory $\alpha$ to a Euclidean curve on the tangent space $T_{\alpha(0)}(\s^2)$. 
What is the space in which these curves lie?
For any point $p \in \s^2$, we denote the set of square-integrable Euclidean curves on the tangent space at $p$ 
as $\cC_p = \ltwo([0, 1], T_p(\s^2))$; $\cC_p$ represents all trajectories on $\s^2$ that start from $p$. 
Therefore, the full space of interest becomes a vector bundle $\cC = \coprod_{p \in \s^2}\ltwo([0, 1], T_p(\s^2))$, 
which is the disjoint union of $\cC_p$ over $p \in \s^2$.
This representation of spherical trajectories is invertible: any trajectory $\alpha$ is uniquely 
represented by a pair $(p,q(\cdot)) \in \cC$, where $p = \alpha(0)$ is the starting point and $q(\cdot) \in \cC_p$ is its TSRVC. 
(We will write $q(\cdot)$ as $q$ to reduce notation.) One can numerically reconstruct the trajectory $\alpha$ 
from its representation $(p,q)$ using Algorithm \ref{algo1}.

\begin{algorithm}\label{algo1}
({\bf Covariant integral of $q$ along $\alpha$})

Given the pair  $(p,q) \in \cC$, we seek a trajectory
$\alpha$ such that $\alpha(0) = p$  and $q \in \cC_p$ is the TSRVC of $\alpha$. 
Suppose $q$ is sampled at $T$ equally-spaced times $\{ t \delta | t=0,1,2, \dots, T-1\}, \delta= 1/T$.
Then, the original trajectory $\alpha$ can be recovered as follows: 
\begin{enumerate}
\item Set $\alpha(0)=p$, and compute $\alpha(\delta)=\exp_{\alpha(0)}(\delta q(0) |q(0)|)$, where $\exp$ denotes the exponential map on $\s^2$ (See Appendix A).
\item For $t=1, 2, \dots, T-1$
\begin{enumerate}
\item Parallel transport $q(t\delta)$ to $\alpha(t\delta)$ along current trajectory from $\alpha(0)$ to $\alpha(t\delta)$, and call it $q^{\parallel}(t\delta)$. 
\item Compute $\alpha((t+1)\delta)=\exp_{\alpha(t\delta)}(\delta  q^{\parallel}(t\delta)   |q^{\parallel}(t\delta)|)$. 
\end{enumerate}
\end{enumerate}
\end{algorithm}
This numerical covariant integration results in a trajectory $\alpha$ whose TSRVC is $q$. 
Next, we develop tools for comparing spherical trajectories, using geodesics and the geodesic distances.

\subsection{Geodesics between Spherical Trajectories} 
The representation space $\cC$ is an infinite-dimensional vector bundle and to 
define geodesic distances on $\cC$, we need to impose a Riemannian structure on it. 
We start by specifying the tangent spaces of $\cC$. 
For an element $(x,v)$ in $\cC$, where $x \in \s^2$ and $v \in \cC_x$, 
we naturally identify the tangent space at $(x,v)$ to be $T_{(x,v)}(\cC) \cong T_x(\s^2) \oplus \cC_x$. 
To see this, suppose we have a path in $\cC$ given by $(\beta(s),q(s,t))$, where 
the path parameter $s \in (-\epsilon,\epsilon), t \in [0,1]$ for a small $\epsilon >0$. 
Assume that the path passes through $(x,v)$ at time $0$, 
i.e. $\beta(0)=x$ and $q(0,t) = v(t)$ for $t \in [0,1]$. 
Note that this path has two components: a baseline $\beta(\cdot)$, which is a curve on $\s^2$, 
and $q(s,\cdot)$ which is a Euclidean curve in $\cC_{\beta(s)}$ for each $s$. The velocity vector of this path at $s=0$ is given by $(x_s(0),\nabla_{x_s(0)}q(0,\cdot)) \in T_x(\s^2) \oplus \cC_x$, where $x_s$ denotes $dx/ds$, and $\nabla_{x_s}$ denotes 
covariant differentiation of tangent vectors.

Now, the Riemannian inner product on $\cC$ is defined in a natural way: 
If $(u_1,w_1)$ and $(u_2,w_2)$ are both elements of $T_{(x,v)}(\cC) \cong T_x(\s^2) \oplus \cC_x$, define
\begin{equation}{\label{eqn:inner}}
\left<(u_1,w_1),(u_2,w_2)\right> = (u_1 \cdot u_2) + \int_0^1(w_1\cdot w_2)~dt \quad,
\end{equation}
where the ``dot'' products in the right indicate the original Riemannian inner product defined on $T_x(\s^2)$. 
The next challenge is to find geodesics between arbitrary  points in $\cC$ 
under this Riemannian metric, and the following result characterizes these geodesics. 
\begin{prop}
If a path $(\beta(s),q(s,\cdot))$, $s\in[0,1]$ is a geodesic in the vector bundle $\cC$, then it has the following properties: 
\begin{enumerate}
\item The base curve $\beta(\cdot)$ is constant-speed parametrized.  That is, $|\dot{\beta}(s)|$ is constant for all $s \in [0,1]$.
\item The TSRVC part $q(s,\cdot)$ is covariantly linear along $\beta$.  That is, 
$\nabla_{\beta_s}(\nabla_{\beta_s} q(s,t)) = 0$ for all $t,s\in[0,1]$. 
\end{enumerate}
\end{prop}
We sketch a proof of this proposition in Appendix \ref{app:twopt} for more details. It is important to note that
inverse does not hold, i.e.  these properties do not imply that the path is a geodesic. 
For any smooth path $(\beta(s),q(s,\cdot))$ that connects two points $(p_1,q_1)$ and $(p_2,q_2)$ on $\cC$ with $(\beta(0),q(0,\cdot))= (p_1,q_1)$ and $(\beta(1),q(1,\cdot))= (p_2,q_2)$ and satisfies these two properties, 
its path length under chosen metric (\ref{eqn:inner}) is given by:
\begin{equation} \label{eqn:pathlen}
 l_\cC({(\beta,q)}) = \sqrt{l_{\beta}^2 + \int_0^1\|{q}^{\parallel}_{1,\ \beta}(t)- q_2(t)\|^2dt} \quad ,
\end{equation}
where $l_{\beta}^2$ represents the squared length of $\beta$ on $\s^2$, defined by $\int_0^1|\dot{\beta}(s)|^2ds$,  and $q^{\parallel}_{1,\beta} $ represents $\left(q_1\right)_{\beta(0)\rightarrow\beta(1)}$, the parallel transport of $q_1$ from $T_{\beta(0)}(\s^2)$ to $T_{\beta(1)}(\s^2)$ along $\beta$ on $\s^2$ (the same space where $q_2$ lies).

As mentioned above, there may be many paths connecting $(p_1,q_1)$ and $(p_2,q_2)$ that satisfy the above two properties, 
but are not geodesics.
The geodesic, by definition, is the shortest one among those. In the following, we layout a way to identify the geodesic path. 
If a path satisfies the two properties listed above, 
it is completely determined by the baseline $\beta$. 
The reason is that  given $(p_1,q_1), (p_2,q_2) \in \cC$ and the baseline $\beta$, the choice of 
$q(s,t)$ is restricted due to the second property (covariant linearity). Therefore, to find the geodesic, the key is to find the optimal baseline $\beta^*$ that minimizes the length of $(\beta(s),q(s,\cdot))$, $s\in[0,1]$ defined in (\ref{eqn:pathlen}):
\begin{equation}{\label{optbeta}}
{\beta^*}=\argmin_{\beta \in {\cal B}}\  \left( \ l_{\beta}^2  + \int_0^1\|{q}^{\parallel}_{1,\ \beta}(t)- q_2(t)\|^2dt  \right) \quad,
\end{equation}
where ${\cal B}$ denotes the space of all valid paths. At this stage, we have ${\cal B}=\{\beta:[0,1]\rightarrow \s^2| \beta(0) = p_1, \beta(1) = p_2 , | \dot{\beta}(s)| \text{ is constant} \}$, which is still a large set. 
To reduce the size of ${\cal B}$, we define the concept of a {\it p-optimal} (optimal parallel transport) curve on $\s^2$. 

\begin{defn}\label{def:poptimal} ({\bf $p$-Optimality})
Let ${\cal T}: T_{p_1}(\s^2) \mapsto T_{p_2}(\s^2)$ be a 
linear, isometric map between the two vector spaces. 
We define a curve $\beta(t)$, $t\in[0,1]$, on $\s^2$ from $p_1$ to $p_2$, to be $p$-optimal if: 
(1) the  parallel transport map induced by the path $\beta$ from $p_1$ to $p_2$ is the same as ${\cal T}$, and 
(2) $\beta$ is  the shortest amongst all such curves.
\end{defn}
Consider a geodesic path connecting $p_1$ and $p_2$ on $\s^2$. This geodesic is a $p$-optimal 
curve if ${\cal T}$ is set to the mapping induced by parallel transport from 
$p_1$ to $p_2$ (along the geodesic). With this definition of $p$-optimal 
curves, we have the following lemma.
\begin{lemma}\label{lem:geodcc}
 If a path $(\beta^*(s),q(s,\cdot))$, $s \in [0,1]$, on $\cC$ is a geodesic, the baseline $\beta^*$ is a p-optimal curve.  
\end{lemma}
The proof of this lemma is simple. If $(\beta^*(s),q(s,\cdot))$, $s \in [0,1]$ is a geodesic path connecting $(p_1,q_1)$ and $(p_2,q_2)$ on $\cC$,
then it will have the shortest length. So if we can find a shorter curve than $\hat{\beta}^{*}$ from $p_1$ to $p_2$, that induces the same parallel transport map, then we could reduce $l_{\beta^*}^2 $ without affecting $\int_0^1\|{q}^{\parallel}_{1,\ \beta^*}(t)- q_2(t)\|^2dt$. This implied that $(\beta^*(\cdot),q(\cdot,\cdot))$ would not be a geodesic on $\cC$.

Lemma 1 helps us reduce the search space for the desired geodesic to a smaller set 
${\cal B}=\{\beta:[0,1]\rightarrow \s^2| \beta(0) = p_1, \beta(1) = p_2 , \beta \text{ is a $p$-optimal curve and } | \dot{\beta}(s)| \text{ is constant} \}$. 
The next result further characterizes elements of this set. 
\begin{lemma}\label{lem:arc}
For any two points $p_1, p_2 \in \s^2$,  the only $p$-optimal curves connecting them are the circular arcs between them.
\end{lemma} 
See Appendix \ref{app1} for the proof. 
Using Lemma \ref{lem:geodcc} and Lemma \ref{lem:arc}, the optimization in Eqn. \ref{optbeta} can now be
highly simplified: the $\argmin$ can be taken over all the circular arcs connecting $p_1$ and $p_2$: ${\cal B}=\{\beta:[0,1]\rightarrow \s^2| \beta(0) = p_1, \beta(1) = p_2 , \beta \text{ is a circular arc} \}$. We propose the following method in Algorithm \ref{algo2} to generate all circular arcs from two points $p_1$ to $p_2$ in $\s^2$. 

\begin{algorithm}\label{algo2}
({\bf Generating circular arcs from $p_1$ to $p_2$}. )

Generate a unit vector $v_1 \in T_{p_1}(\s^2)$ and 
a unit vector $v_2 \in T_{p_2}(\s^2)$.
For each $\theta \in [0, 2 \pi)$, 
\begin{enumerate}
\item Compute the rotation matrix, for rotation by angle $\theta$ about an axis  $p_2$, using the formula:
$R=I \cos \theta+\sin \theta\ [p_2]_\times+(1-\cos \theta)\ p_2 \otimes p_2\ \ ,$
where $[p_2]_\times$ is the cross product matrix of $p_2$, $\otimes$ is the tensor product and $I$ is the identity matrix. This is a matrix form of Rodrigues' rotation formula. 
\item Compute two frames $f_1=[p_1, v_1, w_1]$ and $f_2=[p_2, R v_2, w_2]$, where $w_1$ is the cross product of $p_1$ and $v_1$ and
          $w_2$ is the cross product of $p_2$ and $Rv_2$. 
\item Generate ${\beta}_{\theta}(s)=e^{s A_{\theta}} p_1, s \in [0,1], \mbox {where}\ A_{\theta}=\mbox {logm}\ (f_2 f^T_1) \ .$
\end{enumerate}
\end{algorithm}
In this way, one can generate a one-parameter family of circular arcs $\beta_{\theta}$, connecting $p_1$ and $p_2$ and 
indexed by $\theta \in [0,2\pi)$. 
The optimization problem in (\ref{optbeta}) now becomes: 
\begin{equation}\label{GridSearch}
\theta^*=\argmin_{\theta \in [0, 2\pi)}  \left( l_{\beta_\theta}^2 + \int_0^1\|{q}^{\parallel}_{1,\ \beta_\theta}(t)- q_2(t)\|^2dt  \right)\ .
\end{equation}
We 
will call the resulting optimal curve $\beta^* = e^{sA_{\theta^*}}p_1$, obtained using Algorithm \ref{algo2} and 
an exhaustive search over $\theta$.

After having $\beta^*$,the optimal baseline curve,  the desired geodesic path in $\cC$
can be written as $(\beta^*(s),q(s,t))$, $s,t\in[0,1]$, where $q(s,t)$ is covariantly linear and $q(0,t) = q_1(t)$, $q(1,t) = q_2(t)$.  
More precisely, $q(s,\cdot) = \left((q_{1} + s w_1) \right)_{\beta^*(0) \rightarrow \beta^*(s)}$, where $w_1$ denotes the difference of $q_1$ and $q_2$ at $T_{p_1}(\s^2)$ (defined as $w_1 = (q_2)_{\beta^*(1) \rightarrow \beta^*(0)} - q_1 $. According to (\ref{eqn:pathlen}), the length of the geodesic path is given by:
\begin{equation}\label{geodistance}
d((p_1,q_1), (p_2,q_2))=\sqrt{ l_{\beta^*}^2 + \int_0^1\|{q}^{\parallel}_{1,\beta^*}(t)- q_2(t)\|^2dt } \ \ .
\end{equation}

\begin{figure}
\begin{center}
\begin{tabular}{ccc}
\includegraphics[height=1.4in]{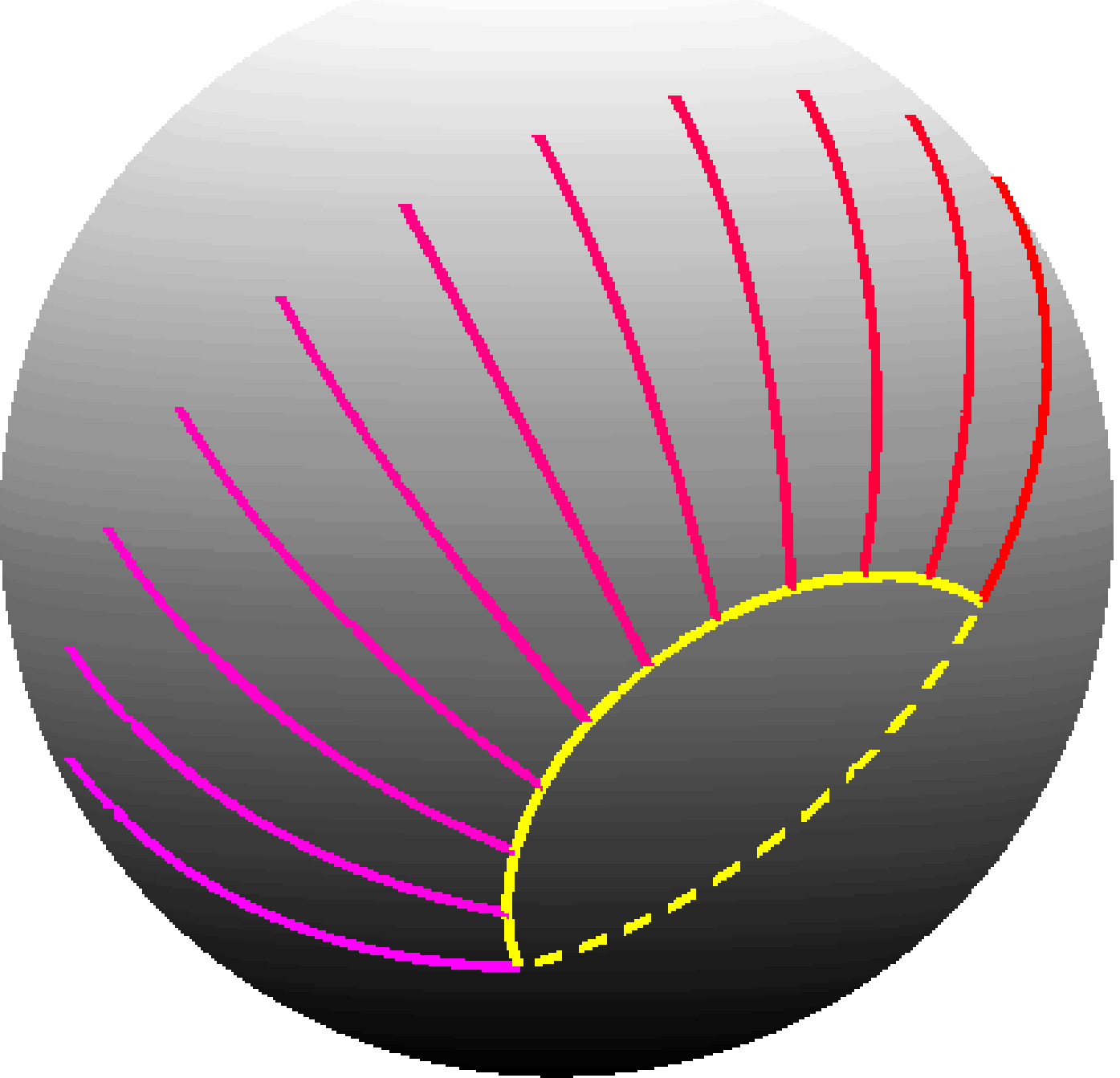}&
\includegraphics[height=1.4in]{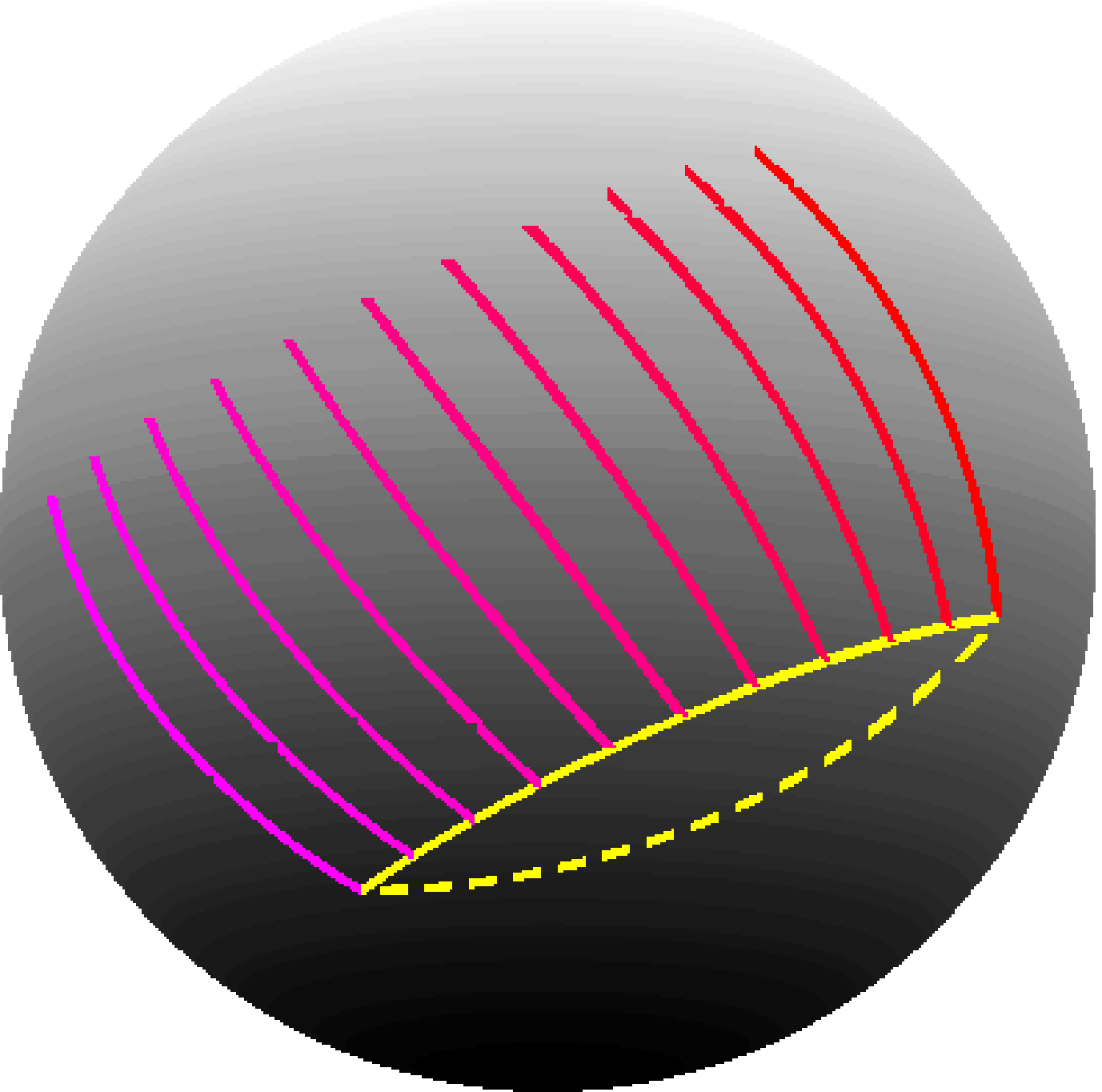}&
\includegraphics[height=1.4in]{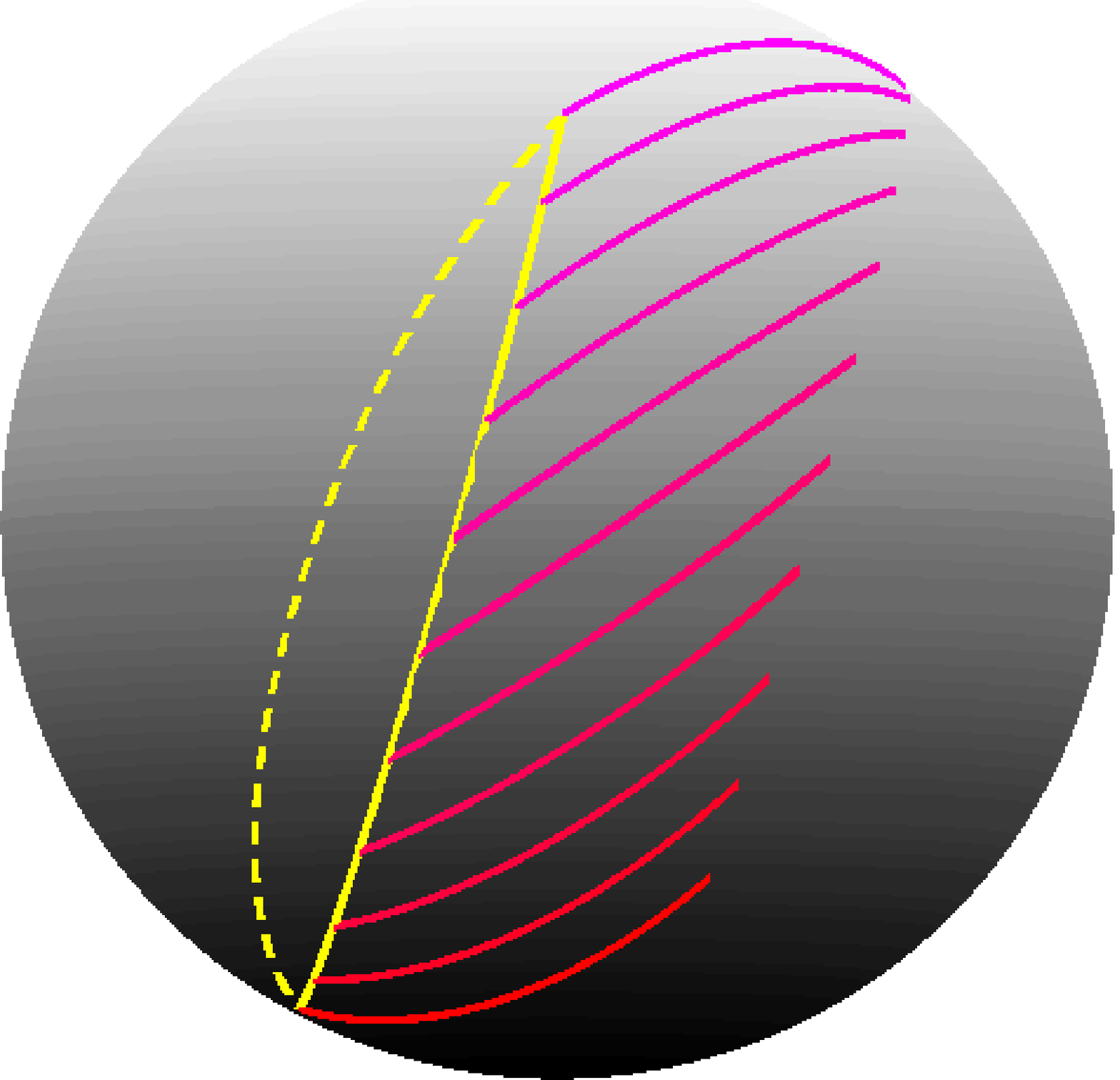}
\\
\end{tabular}
\caption{Example of geodesics between two trajectories in $\s^2$. The $p$-optimal baseline trajectory is 
shown in the solid line while a simple $\s^2$-geodesic between the end points is shown in the 
broken line.}\label{ill:trajgeod} 
\end{center}
\end{figure}

For displaying  a geodesic path, we can recompute the trajectories on $\s^2$ for each $s \in [0,1]$, 
using the numerical covariant integral laid out in Algorithm \ref{algo1}. 
That is, map  $(\beta^*(s), q(s,\cdot)$ back to a trajectory on $\s^2$ by treating  $\beta^*(s)$ 
as the starting point and $q(s,\cdot)$ as its TSRVC. Fig. \ref{ill:trajgeod} shows three examples of 
geodesic paths. In each case,  
the yellow solid line shows the optimal baseline $\beta^*$ and 
the yellow dash line shows the simple $\s^2$-geodesic connecting $\beta^*(0)$ and $\beta^*(1)$ on $\s^2$ for comparison.

\subsection{Phase-Amplitude Separation of Two Trajectories}
The main motivation for using TSRVC representation comes from the following theorem.
If a trajectory $\alpha$ is warped by $\gamma$, resulting in $\alpha \circ \gamma$, what is the TSRVC
of the time-warped trajectory? This TSRVC is given by:
\begin{eqnarray*}
q_{\alpha \circ \gamma}(t) &=& {  \left( (\dot{\alpha}(\gamma(t)) \dot{\gamma}(t)) \over \sqrt{ |  \dot{\alpha}(\gamma(t)) \dot{\gamma}(t)|} \right)}_{\alpha(\gamma(t)) \rightarrow p} =
{ \left( (\dot{\alpha}(\gamma(t))) \sqrt{\dot{\gamma}(t)}  \over \sqrt{ |  \dot{\alpha}(\gamma(t)) |} \right)_{\alpha(\gamma(t)) \rightarrow p}} \\
&=& q_\alpha(\gamma(t)) \sqrt{\dot{\gamma}(t)} \equiv (q_\alpha* \gamma)(t)\ .
\end{eqnarray*}
Under the TSRVC representation, the action of time-warping of original trajectories under the metric $d$ in  (\ref{geodistance}) is by isometries. 
\begin{thm} \label{thm:iso}
For any two trajectories $\alpha_1, \alpha_2 \in {\cal F}$, and the corresponding representation $(p_1,q_1), (p_2,q_2) \in \cC$, the metric
$d$ given in Eqn. \ref{geodistance} 
satisfies $d((p_1,q_{\alpha_1\circ \gamma}), (p_2,q_{\alpha_2 \circ \gamma})) = d((p_1,q_{\alpha)1}), (p_2,q_{\alpha_2}))$, for any $\gamma \in \Gamma$.
\end{thm}
The proof of this theorem is presented in Appendix \ref{app2}. 
This property termed the {\it isometry} of (time-warping) action under the metric $d$, 
allows us to fully perform phase-invariant comparisons and analysis of trajectories. 
This is achieved by defining a metric in the amplitude space of trajectories.

To formally define the amplitude of a trajectory, we introduce $\tilde{\Gamma}$, 
the set of all non-decreasing, absolutely continuous functions $\gamma$ on $[0,1]$ such that $\gamma(0) = 0$ and $\gamma(1) = 1$. 
$\tilde{\Gamma}$ is a larger set of time-warping functions than $\Gamma$ and
it can be shown that $\Gamma$ is a dense subset of $\tilde{\Gamma}$ \citep{su2014}.

\begin{defn}({\bf Trajectory Amplitude}) For any trajectory $\alpha$, we define its amplitude to be the set of all 
possible time warpings of $\alpha$ under $\tilde{\Gamma}$. In the representation space $\cC$,  
this amplitude corresponds to the set: 
$$ [(p,q)] \equiv (p,[q]) = \{(p,(q\circ\gamma)\sqrt{\dot{\gamma}})|\gamma \in \tilde{\Gamma} \}. $$
\end{defn}
We define the amplitude space ${\cal A}$ as the set of amplitude of all trajectories.

Each amplitude is considered as an equivalence class. Any two trajectories $\alpha_1, \alpha_2$,
with representations of $(p_1,q_1)$ and $(p_2,q_2)$, 
are deemed equivalent if: (1) $p_1 = p_2$; and (2) there exist a sequence $\gamma_i \in \tilde{\Gamma}$ 
such that $q_{\alpha_2 \circ \gamma_i}$ converges to $q_1$. 
Theorem \ref{thm:iso} indicates that if two trajectories are warped by the same $\gamma$ function, 
then the distance $d$ between them remains the same. This leads to the definition of an amplitude distance 
between trajectories. 

\begin{defn}({\bf Ampltude Distance})
For any two amplitudes $(p_1,[q_1])$ and $(p_2,[q_2])$ in ${\cal A}$, the amplitude distance between them is defined to be: 
\begin{eqnarray}
\nonumber d_a((p_1,[q_1]),(p_2,[q_2])) &=& \inf_{\gamma_1,\gamma_2 \in \tilde{\Gamma}}d((p_1,(q_1\circ\gamma_1)\sqrt{\dot{\gamma_1}}),(p_2,(q_2\circ\gamma_2)\sqrt{\dot{\gamma_2}})) \\ 
\label{eqn:d_q} & \approx & \inf_{\gamma \in \Gamma} d((p_1,q_1),(p_2,(q_2\circ\gamma)\sqrt{\dot{\gamma}}))\ ,
\end{eqnarray}
where the last approximation comes from the fact that $\Gamma$ is dense in $\tilde{\Gamma}$. 
\end{defn}
As stated earlier, our goal is to separate phase and amplitude of given trajectories, and then to compare them in a
way that is independent of their phases.  The above definition achieves that goal for pairwise comparisons. 
Note that Eqn. \ref{eqn:d_q} provides not only a distance between two amplitudes, which is invariant of the phases
of $\alpha_1$ and $\alpha_2$, 
but it also gives the optimal time-warping function $\gamma$ to align trajectory $\alpha_2$ to $\alpha_1$. 
That is the point $\alpha_1(t)$ is optimally matched with the point $\alpha_2 (\gamma(t))$. 
This $\gamma$
is called the {\it relative phase} of $\alpha_1$ with respect to $\alpha_2$. 

For a fixed $\gamma$, calculating the 
distance $d((p_1,q_1),(p_2,(q_2\circ\gamma)\sqrt{\dot{\gamma}}))$ is an optimization problem with respect to $\theta \in [0,2\pi)$, as discussed earlier. Hence, (\ref{eqn:d_q}) represents a two-parameter optimization problem:
\begin{equation} \label{eqn:two_para}
\min_{\theta, \gamma \in [0,2\pi) \times \Gamma}\left( {\ l_{\beta_\theta}^2 + \int_0^1\|{q}^{\parallel}_{1,\ \beta_\theta}(t)- (q_2* \gamma)(t)\|^2dt\ }     \right)
\end{equation}
To solve the two-parameter optimization problem, we use the following strategy: for each $\theta \in [0,2\pi]$, we optimize over $\gamma \in \tilde{\Gamma}$ and then find the best combination of $\theta$ and $\gamma$. The optimization over $\gamma$ is solved using Dynamic Programming algorithm \citep{BertsekasDP01}. The algorithm is summarized below:
\begin{algorithm}\label{algo4}
{\bf Computation of  Amplitude Distance}
\begin{enumerate}
\item For each $\theta \in [0, 2\pi)$, solve ${\hat \gamma} $ by Dynamic Programming:
$$
{\gamma^*_\theta}=\argmin_{\gamma \in {\Gamma}} \left( \int_0^1\|{q}^{\parallel}_{1,\ \beta_\theta}(t)- (q_2* \gamma(t))\|^2dt \right)\ \,
$$
and let $E(\theta,\gamma^*_\theta) = l_{\beta_\theta}^2 + \int_0^1\|{q}^{\parallel}_{1,\ \beta_\theta}(t)- (q_2* \gamma^*_\theta(t))\|^2dt$

\item Find $({\theta^*}, \gamma^*_{\theta})$ such that ${\theta^*}=\argmin_{\theta \in [0, 2\pi)} \left( E(\theta,\gamma^*_\theta)    \right)$. 
The minimum value of $E$ is the amplitude distance $d_a$. 
\end{enumerate}
\end{algorithm}

\subsection{Phase-Amplitude Separation of Multiple Trajectories} 
Using the amplitude distance,
we can calculate sample mean and modes of variations for given a collection of trajectories, 
while being invariant to their phases. 
The mean trajectory can be treated as a template for registering the set of trajectories, 
i.e. phase-amplitude separation. In this paper, the sample mean is calculated through the notion of Karcher mean \citep{CPA:Karcher}
under $d_a$.   Given a set of trajectories $\{\alpha_i, \ i= 1\dots n\}$, and their representations $\{ (p_i,q_i), i = 1\dots n \}$, their Karcher mean in the amplitude space ${\cal A}$ is defined to be: 

\begin{equation} \label{eqn:karchermean}
(\mu_p,[\mu_q]) = \argmin_{(p,[q]) \in {\cal A}} \sum_{i=1}^n d_a((p,[q]),(p_i,[q_i]))^2\ .
\end{equation}
Note that $(\mu_p,[\mu_q])$ is an orbit (equivalence class) and one can select any element of this orbit as a template to help to align multiple trajectories. 
Since the original space $\cC$ is a nonlinear Riemannian manifold, the optimization of  (\ref{eqn:karchermean}) requires the {\it exponential map} and {\it inverse exponential map}. In the following, we will define the {\it exponential map} and {\it inverse exponential map} on $\cC$ .

\noindent {\bf Inverse Exponential Map on $\cC$}: Given $(p_1, q_1)$ and $(p_2, q_2)$, let $\{(\beta(s), q(s,\cdot)) | s\in [0,1]\}$ be the geodesic connecting them on $\cC$.  The inverse exponential map from $(p_2,q_2)$ to $(p_1,q_1)$ is defined to be the mapping from 
$\cC$ to $T_{(p_1, q_1)}(\cC)$ such that 
$\exp^{-1}_{(p_1, q_1)}(p_2, q_2) = (u_1, w_1)$,
where $u_1 \in T_{p_1}(\s^2)$ and $u_1 \perp p_1$ with $\|u_1\|=l_{\beta}$.
The expressions for $u_1$ and $w_1$  are given below:
\begin{itemize}
\item $u_1=l_{\beta}   { A_{\theta} p_1 \over \| A_{\theta} p_1 \| }$, where $A_{\theta}$ is defined in Algorithm \ref{algo2} such that $\beta(s) = e^{sA_{\theta}}p_1$. It is easy to show that $A_{\theta}$ is an asymmetric matrix, and $trace(A_\theta) = 0$. 
 
\item $w_1 = q^{\parallel}_{2,\beta} - q_1 \in \ltwo ([0, 1], T_{p_1}(\s^2))$, where $q^{\parallel}_{2,\beta}$ denotes the backward parallel transport of $q_2$ along $\beta$ from $p_2$ to $p_1$. 
\end{itemize}

\noindent{\bf Exponential Map on $\cC$}: Given a point $(p_1, q_1)$ in $\cC$ and a tangent vector 
$(u_1, w_1) \in T_{(p_1,q_1)}(\cC)$, the exponential map is a mapping from $T_{(p_1, q_1)}(\cC)$ 
to $\cC$. Furthermore, 
$\exp_{(p_1,q_1)}(s(u_1,w_1))$, with $s \in [0,1]$, 
provides a geodesic path on $\cC$; we will denote it by $(\beta(s), q(s,\cdot))$. 
Since the geodesic is determined by the baseline $\beta$,  we only need to find an asymmetric matrix $A$ as described in Algorithm \ref{algo2} such that $e^{sA}p_1 = \beta(s)$ to determine the baseline. Here, $A$ has three unknown parameters because $trace(A) = 0$. To solve for $A$, we have the following equations:
\begin{enumerate}
\item The vector fields $\dot{\beta}(s)$ take the value $u_1$ at $s=0$ by definition, and we have $ \dot{\beta}(s)|_{s = 0 } =Ap_1$. Therefore, the first equation for solving $A$ is
\begin{equation} 
Ap_1= u_1.
\label{eqn:exp1}
\end{equation}

\item According to the geodesic equations derived 
in \citet{Zhang2015} on $\cC$, the second derivative of the baseline $\beta$ is determined by: 

$$\nabla_{\dot{\beta}(s)}(\dot{\beta}(s)) = -R(q(s),(\nabla_{\dot{\beta}(s)}q)(s)) \dot{\beta}(s), $$
where $R(\cdot,\cdot)$ denotes the Riemannian curvature tensor. In the left side, we have $\nabla_{\dot{\beta}(s)}(\dot{\beta}(s))|_{s =0} = P_{T_{p_1}(\s^2)}\left(\frac{d^2}{ds^2} e^{sA}p_1 |_{s=0} \right)$, where $ P_{T_{p_1}(\s^2)}(\cdot)$ denotes the projection of a vector to the tangent space $T_{p_1}(\s^2)$.  In the right side, given the baseline $\beta$, we know that $q(s,\cdot) = \left(q_1+ s w_1\right)_{p_1 \rightarrow \beta(s)} =(q_1)_{p_1 \rightarrow \beta(s)} + s (w_1)_{p_1 \rightarrow \beta(s)}$. Therefore, we have $(\nabla_{\dot{\beta}(s)}q)(s) = (w_1)_{p_1 \rightarrow \beta(s)}$. At the point $s =0$, $R(q(s),(\nabla_{\dot{\beta}(s)}q)(s))\dot{\beta}(s)|_{s=0} =R(q_1, w_1) \dot{\beta}(0)$. Therefore, the above equation simplifies at $s=0$ to become:
\begin{equation} 
P_{T_{p_1}(\s^2)} \left(  \frac{d^2}{ds^2}  e^{sA} p_1 \right) = - R(q_1, w_1) u_1\ .
\label{eqn:exp2}
\end{equation}

\end{enumerate}

(\ref{eqn:exp1}) and (\ref{eqn:exp2}) are used to solve for the asymmetric matrix $A$ as the function of $p_1,q_1,u_1$ and $w_1$.  
(\ref{eqn:exp1}) provides two equations to solve unknown parameters in $A$ since $u_1\perp p_1 $. It is the same for (\ref{eqn:exp2}) which provides two additional equations. So there are four equations for three unknown parameters in $A$ and one is redundant.   Given $A$, the exponential map 
can be expressed as $\exp_{(p_1,q_1)} (s(u_1,w_1)) = (e^{sA}p_1, (q_1 + sw_1)_{p_1 \rightarrow e^{sA}p_1})$.

Once we have specified the exponential and the inverse exponential maps, we can adapt a standard algorithm to find the mean of multiple trajectories 
$\{\alpha_1,\alpha_2,...,\alpha_n \}$. 

%
%
\begin{algorithm}\label{algo5}
{\bf Karcher Mean of Amplitudes} \\
Let $(p_i,q_i)$ denote the pair representation of the trajectory $\alpha_i$, where $p_i = \alpha_i(0)$ and  $q_i$ is its TSRVC.  Let $(\mu_p^j,\mu_q^j), j = 0$ be the initial estimate of the Karcher mean.
\begin{enumerate}

\item For each $i=1,...,n$, align each trajectory $(p_i,q_i)$ to $(\mu_p^j,\mu_q^j)$ according to Eqn. (\ref{eqn:two_para}) using Algorithm \ref{algo4}, and let $\theta^*_i$ and $\gamma_{i}^*$ denote the optimal solution. The aligned trajectory is given as $\tilde{\alpha}_i = \alpha_i \circ \gamma^*_{i}$, and its representation is denoted as $(p_i,\tilde{q}_i)$, where $\tilde{q}_i = (q_i*\gamma^*_i)$.

\item Compute the inverse exponential map: $(u_i, w_i) = \exp^{-1}_{(\mu_p^j,\mu_q^j)}(p_i, \tilde{q}_i)$, where $u_i = l_{\beta_{\theta_i^*}}\frac{A_{\theta^*_i}u_p^j}{\|A_{\theta^*_i}u_p^j\|}$, and $w_i = \tilde{q}_i - u_q^j $.

\item Compute the average direction: $\bar{u} = \frac{1}{n}\sum_{i=1}^n u_i$, $\bar{w} = \frac{1}{n}\sum_{i=1}^n w_i$.

\item If $||\bar{u}|| + || \bar{w} ||$ is small, stop. Otherwise, update $(\mu_p^j,\mu_q^j)$ in the direction of $(\bar{u},\bar{w})$ according to
$$ (\mu_p^{j+1},\mu_q^{j+1}) = \exp_{(\mu_p^j,\mu_q^j)}(\epsilon (\bar{u}, \bar{w})), $$ 
where $\epsilon$ is a small step size, typically $0.5$.

\item Set $j=j+1$, return to step 1. 
\end{enumerate}
\end{algorithm}
For the final output $(\mu_p,\mu_q)$, we can reconstruct the mean trajectory using Algorithm \ref{algo1}, 
denoted by $\mu$. Note that Algorithm \ref{algo5} provides three sets of output: (1) the mean trajectory $\mu$; (2) the aligned trajectories $\{\tilde{\alpha}_i\}$ representing the {\it amplitudes}; and (3) the warping-functions or the phase components  $\{\gamma^*_i\}$. 

\section{Statistical Modeling of Amplitudes of Trajectories}
In this section, we use this framework to discover essential modes of 
variability in spherical trajectories and to 
develop statistical models for capturing variability in amplitudes of these trajectories. 

\subsection{Analysis of Modes of  Amplitude Variability} \label{sec:pca}
First, we consider the problem of discovering 
dominant modes of variability in a training data. This is 
achieved using manifold functional PCA (mfPCA), as described below. 
As a pre-processing step,  assume that we have extracted the amplitude components 
 $\tilde{\alpha}_1, \cdots \tilde{\alpha}_n$ of a given set of trajectories $\alpha_1, \cdots, \alpha_n$, 
 using Algorithm \ref{algo5}.

Let $(p_1, \tilde{q}_1),\cdots, (p_n, \tilde{q}_n)$ and $(\mu_p,\mu_q)$ be the representations of the 
aligned trajectories and the mean, respectively, in $\cC$.  
The main difficulty in performing mfPCA is the nonlinearity of $\cC$. To overcome this problem, we choose the tangent space 
$T_{(\mu_p,\mu_q)}(\cC)$, a vector space given by $T_{\mu_p}(\s^2) \oplus \cC_{\mu_p}$, as the setting for PCA. 
The outcomes of Algorithm \ref{algo5} include the Karcher mean $(\mu_p,\mu_q)$ and the shooting vectors 
(also the tangent vectors) $(u_i,w_i)$, associated with the amplitudes $(p_1, \tilde{q}_1),\cdots, (p_n, \tilde{q}_n)$ on $T_{(\mu_p,\mu_q)}(\cC)$. Each shooting vector has by two parts: $u_i \in T_{\mu_p}(\s^2)$, a vector of $\real^3$, and $w_i \in \cC_{\mu_p}$, an $\ltwo$ function on $T_{\mu}(\s^2)$. Note that $T_{\mu_p}(\s^2)$ is a two-dimensional space. Therefore, we define a new coordinate system using two orthogonal unit vectors $v_1, v_2 \in \real^3$ on  $T_{\mu_p}(\s^2)$, and use the new coordinate system to represent each vector $u_i$ and each function $w_i$. Under this new coordinate system, $u_i$ is a vector in $\real^2$ and $w_i$ is a function in $\ltwo([0,1],\real^2)$. 

 For mfPCA, we will treat the two parts in the shooting vector $(u_i,w_i)$ separately, by computing 
 separate covariance matrices: (1) a sample covariance matrix for $\{u_i\}$,  
 ${\bf K}_{u} = \frac{1}{n-1}\sum_{i=1}^n u_i u_i^t$; and (2) a sample covariance function for 
 $\{w_i\}$, ${\bf K}_w(t_1,t_2)\rightarrow \frac{1}{n-1}\sum_{i=1}^n \left<w_i(t_1),w_i(t_2)\right>$. 
 In practice, each function $w_i$ is sampled at a finite number of points, say $T$, and the resulting covariance function is 
 stored as a matrix.  In most cases, the observation size $n$ is much less than $T$ and, consequently, $n$ controls the degree of variability in the stochastic model. 
Let ${\bf w} \in \real ^{2T\times n}$ be the shooting vectors associated with the TSRVCs of aligned trajectories in $\cC_{\mu_p}$, 
and let ${\bf K}_w \in \real^{2T \times 2T}$ be the sample covariance matrix with  ${\bf K}_w ={ \bf U \Sigma}_w {\bf U}^T$  
as its singular value decomposition (SVD). The submatrix formed by the first $r$ columns of ${\bf U}$, denoted as ${\bf U}_r$, spans the principal subspace of the observed data. The principal coefficients for observations ${\bf w}$ is given as ${\bf C} = {\bf U}_r^T{\bf w} \in \real^{r\times n}$.
 
To visualize the dominant modes of variations, 
we can calculate straight lines along these directions for each component of the shooting vector, and project these lines 
back on $\s^2$ using the exponential map $\exp_{(\mu_p,\mu_q)}(\tau (u,w) )$ for $\tau \in [-1,1]$. 
Here, $u$ is the dominant direction for the location and $w$ is the principal direction for the second component. 

\subsection{Random Sampling from A Wrapped Gaussian Model on Amplitudes} \label{sec:Gmodel}
Our next goal is to impose a simple probability model on the amplitudes of trajectories on $\s^2$, 
and then validate it using random samples from the model. 
There are a few different options for imposing such models  \citep{kurtek2012,mardia2008,anuj2011}.  
We take a common approach where we start with a
probability density in the principal subspace of a tangent space and then map it back to trajectories on $\s^2$ using exponential map. To be more precise, we use the tangent space at the mean $T_{(\mu_p,\mu_q)}(\cC)=T_{\mu_p}(\s^2) \oplus \cC_{\mu_p} $ as the vector space to 
impose a probability model. Since each shooting vector in $T_{\mu_p}(\s^2) \oplus \cC_{\mu_p}$ has two components, we model the two components independently by using multivariate Gaussian models on the principal coefficients. 
For example, for the first component, let ${\bf K}_u = {\bf V \Sigma}_u {\bf V}^T$ be the SVD of the sample covariance for $u_i$, as earlier. A random variable $c_u$ can be sampled from $N({\bf 0},{\bf \Sigma}_u)$, and the corresponding random first component is $u = {\bf V}c_u$. Similarly, for the second component, a random variable $c_w$ can be sampled from $N(0,{\bf \Sigma}_w)$, and the random second component is $w = {\bf U}c_w$. One can reconstruct the sampled trajectory using the exponential map $\exp_{(\mu_p,\mu_q)}((u,w))$ after mapping $u$ and $w$ back to the original coordinate system (using $v_1$ and $v_2$ defined earlier). This provides a technique for sampling from the wrapped Gaussian model on $\cC$.


\section{Experimental Results}
In this section, we present some experimental results to support this elastic framework, 
involving both simulated and real data. These results include computation of 
geodesic paths, computation of mean amplitudes, mfPCA of given spherical trajectories, clustering of 
trajectories under the amplitude distance $d_a$, and random sampling of spherical trajectories under 
a simple statistical model. 

\subsection{Simulated Data}

\begin{figure}[h!]
\begin{center}
\begin{tabular}{|ccc|}
\hline
\multicolumn{3}{|c|}{Example 1}\\
\includegraphics[height=1.6in]{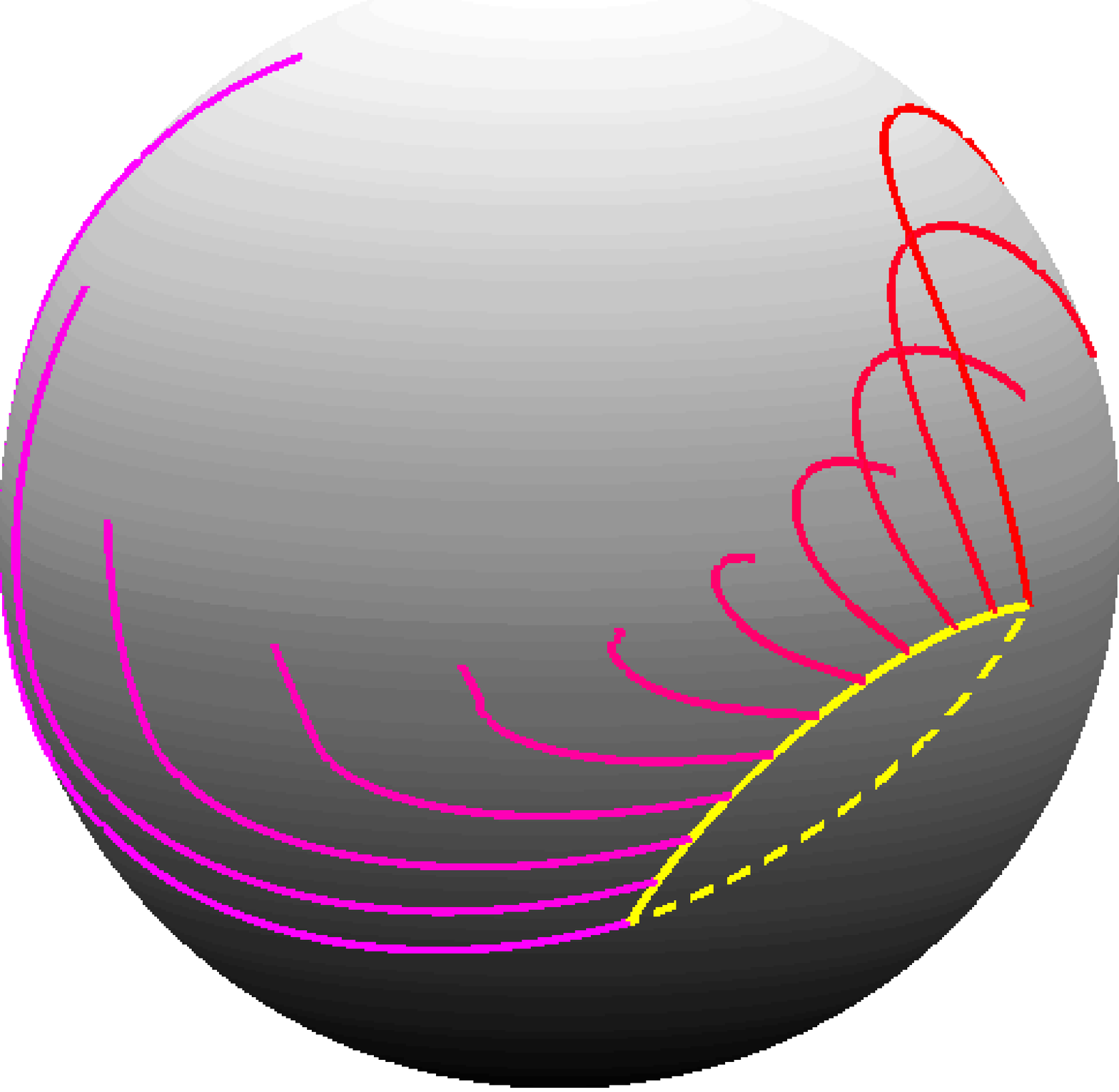}&
\includegraphics[height=1.6in]{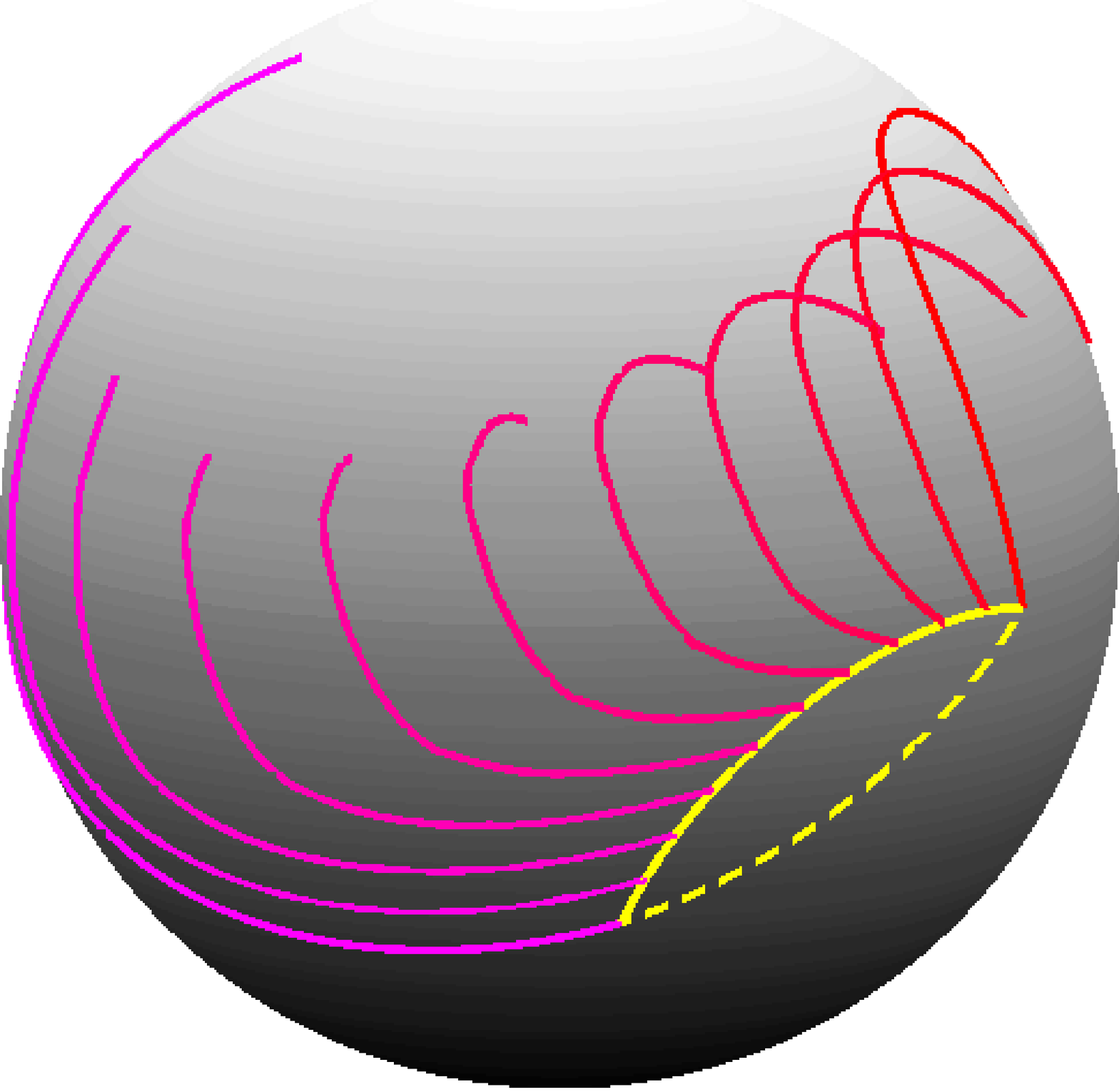}&
\includegraphics[height=1.4in]{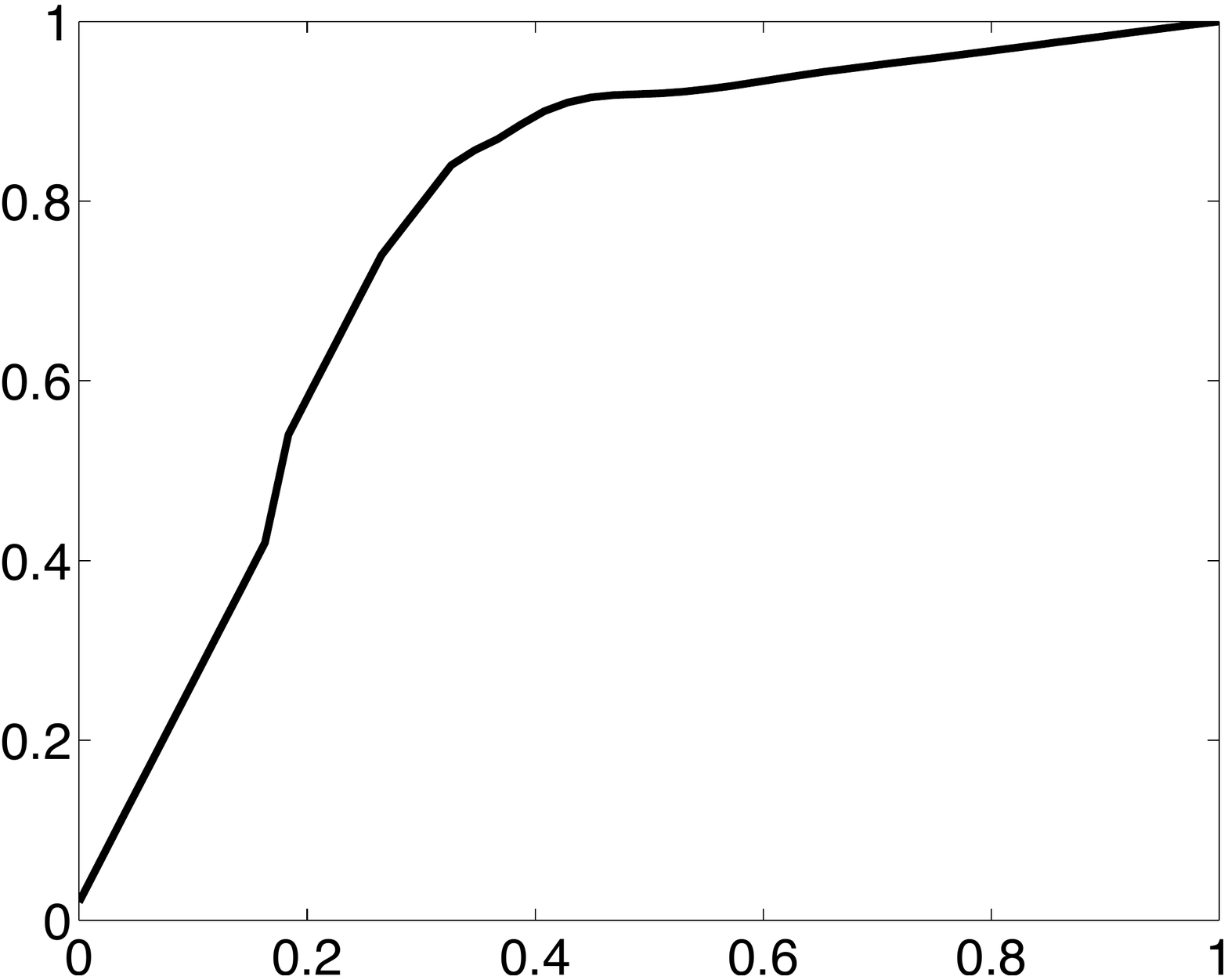} \\
Before registration & After registration & $\gamma^*$ 
\\
\hline
\multicolumn{3}{|c|}{Example 2}\\
\includegraphics[height=1.6in]{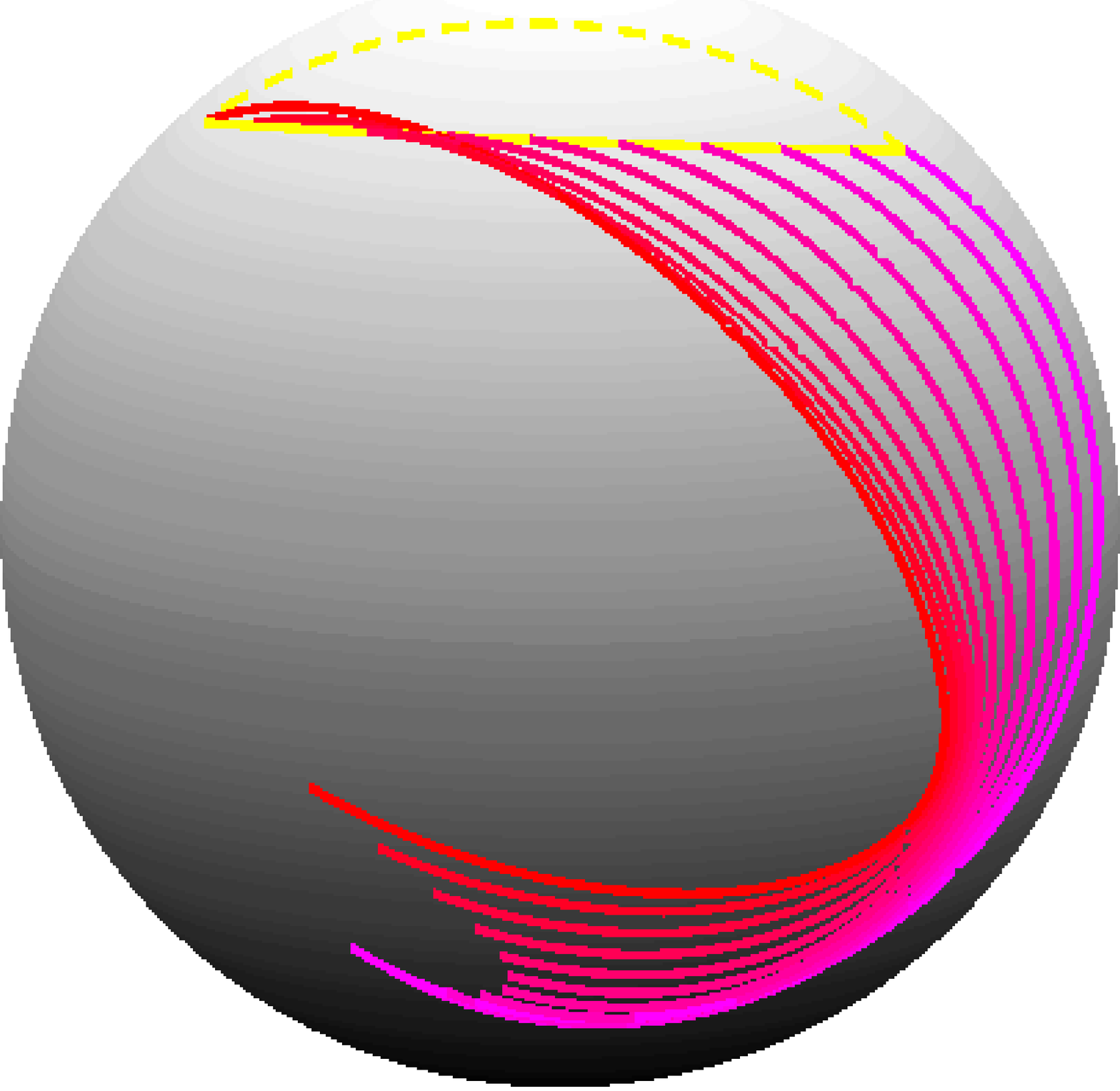}&
\includegraphics[height=1.6in]{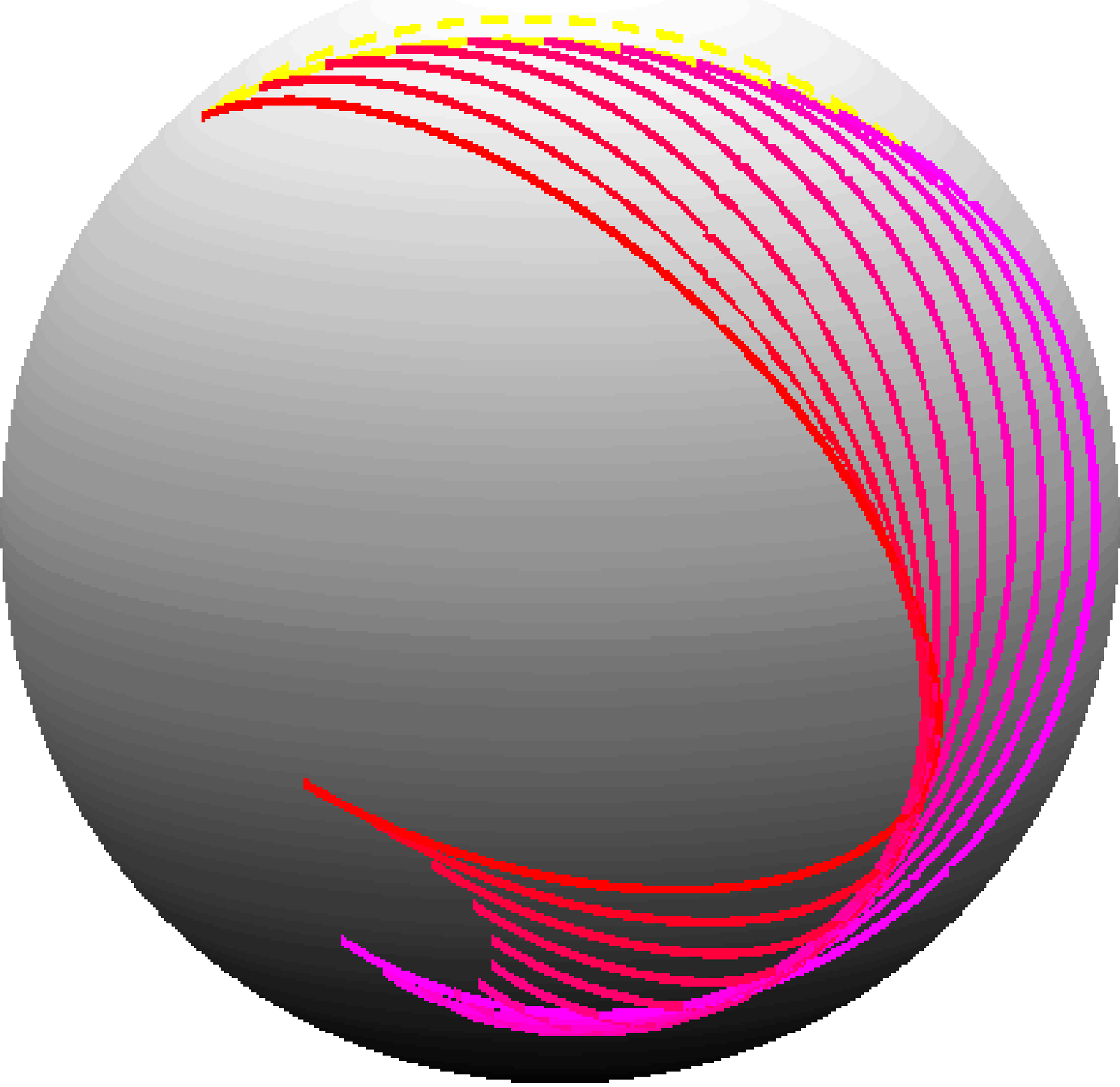}&
\includegraphics[height=1.4in]{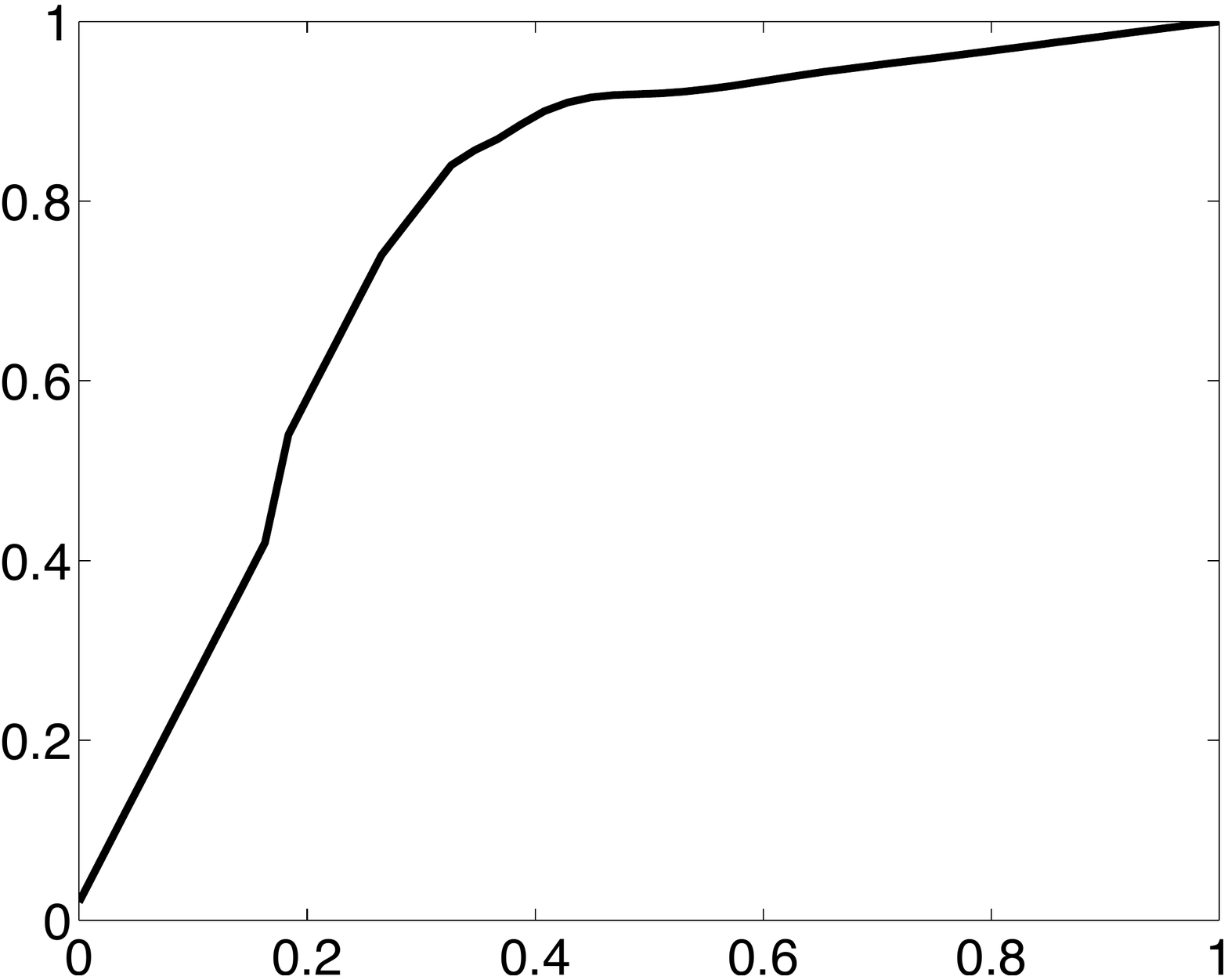}\\
Before registration & After registration & $\gamma^*$ 
\\
\hline
\end{tabular}
\caption{Geodesic paths between spherical trajectories before and after registration. }\label{ArbitraryRegistration} 
\end{center}
\end{figure}

 {\bf Geodesic Computations}: 
To start with we use some simulated spherical trajectories 
and compute geodesic paths between them, {\it without} and {\it with} registration.  
 Fig. \ref{ArbitraryRegistration} shows two examples using arbitrary spherical trajectories. 
In each case,  two corner trajectories form the original given trajectories 
$\alpha_1$ and $\alpha_2$, and the intervening trajectories 
represent equally-spaced sample points along geodesic paths. The optimal baseline curve is denoted by the solid yellow line 
and, for the purpose of comparison, the dashed line denotes the simple $\s^2$-geodesic between the starting points of 
$\alpha_1$ and $\alpha_2$. 
In each example, the first column shows results geodesic without registration, 
the middle column shows results after phase separation and the last column shows the 
relative phase, i.e. the optimal $\gamma^*$ for alignment.   
In both cases, geodesic paths after phase removal better preserve structures during geodesic deformations 
and the resulting distances are much smaller. 
In particular, the elastic geodesic in Example 1 preserves the ``bump"  in the 
trajectory. Also,  the optimal baseline curves noticeably different after phase removal, which further
improves interpretability of geodesic paths. 
After registration, the distance goes from $9.56$ to $d_a = 6.78$ in Example 1, and from $4.41$ to $d_a = 3.05$ in Example 2. 

\begin{figure}[h!]
\begin{center}
\begin{tabular}{|ccc|}
\hline
\multicolumn{3}{|c|}{Example 1}\\
\includegraphics[height=1.2in]{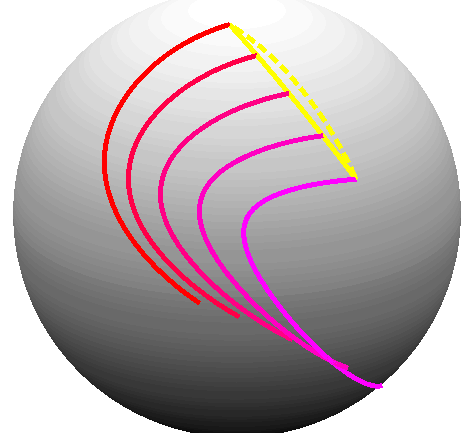}&
\includegraphics[height=1.2in]{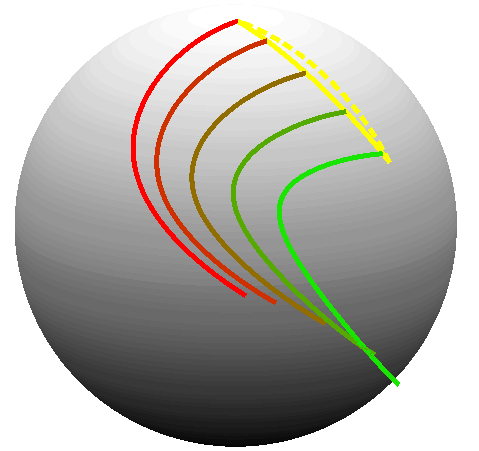}&
\includegraphics[height=1.2in]{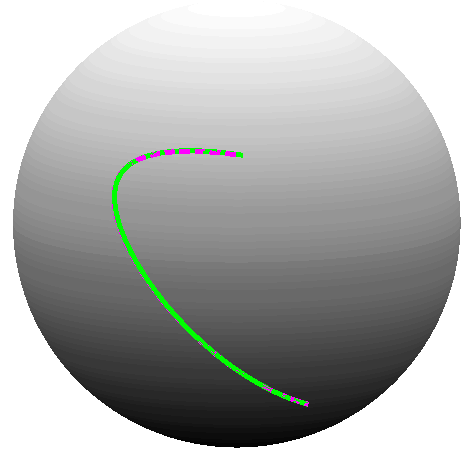}\\
\hline
\multicolumn{3}{|c|}{Example 2}\\
\includegraphics[height=1.2in]{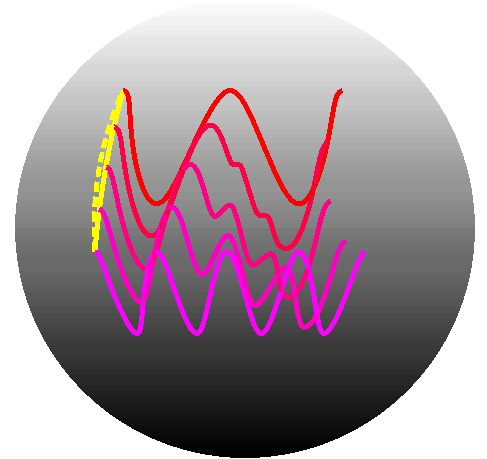}&
\includegraphics[height=1.2in]{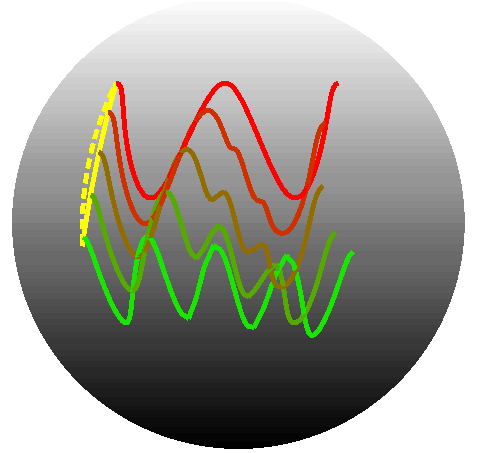}&
\includegraphics[height=1.2in]{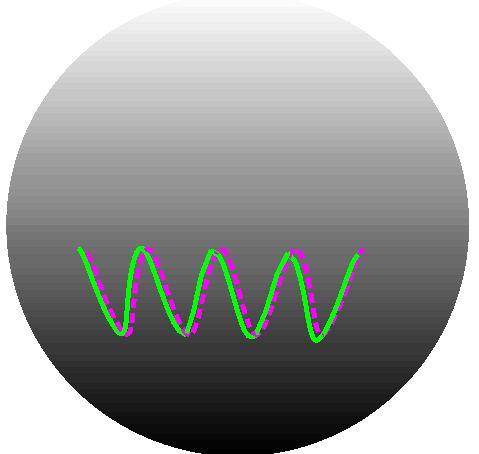}\\
Geodeisc calculated with Alg. \ref{algo4} & Shooting geodeisc & $\exp_{(p_1,q_1)}(u_1,w_1) $ and $ (p_2,q_2)$ \\
\hline
\end{tabular}
\caption{Verification of the exponential map. }\label{shooting} 
\end{center}
\end{figure}

 {\bf Mean Amplitude Computation} or {\bf Phase-Amplitude Separation of Trajectories}:
We can use Algorithm \ref{algo5} to calculate the mean amplitude and to separate the phase and amplitude of given trajectories. 
The mean calculation (Algorithm \ref{algo5})  requires certain computational tools developed in Section 3.4
(exponential map and inverse exponential map on $\cC$), and we first demonstrate their use. 
Once again, given two trajectories $\alpha_1$ and $\alpha_2$, we represent them as $(p_1,q_1)$ and $(p_2,q_2)$, where $p_1,p_2$ are the starting points and $q_1$, $q_2$ are their TSRVCs, respectively. The geodesic between $(p_1,q_1)$ and $(p_2,q_2)$ is calculated using Algorithm \ref{algo4}.  Using the expression for the inverse exponential map, we 
first calculate the shooting vector $(u_1,w_1) = \exp^{-1}_{(p_1,q_1)}(p_2,q_2)$. 
Then, we use the exponential 
 map given by $\exp_{(p_1,q_1)}(s(u_1,w_1)) = (e^{sA}p_1,(q_1+sw_1)_{p_1 \rightarrow e^{sA}p_1})$, where $s\in [0,1]$ and $A$ is solved by (\ref{eqn:exp1}) and (\ref{eqn:exp2}).  Fig. \ref{shooting} shows two examples of the exponential map. In both examples, the first column shows geodesic between two trajectories calculated using Algorithm \ref{algo4} after temporal registration, where the red trajectory is $\alpha_1$ and the pink trajectory is $\alpha_2$.  The second column shows geodesic calculated using exponential map, $\exp_{(p_1,q_1)}(s(u_1,w_1))$. The last column shows $\exp_{(p_1,q_1)}(u_1,w_1)$ (solid line) and the ``target" $\alpha_2$ (dash line).  One can see that the developed exponential map and inverse exponential map works well because the shot path 
 $\exp_{(p_1,q_1)}(u_1,w_1)$ is almost the same as the target trajectory (up to some numerical errors).  

\begin{figure}
\begin{center}
\begin{tabular}{c}
\includegraphics[height=2.6in]{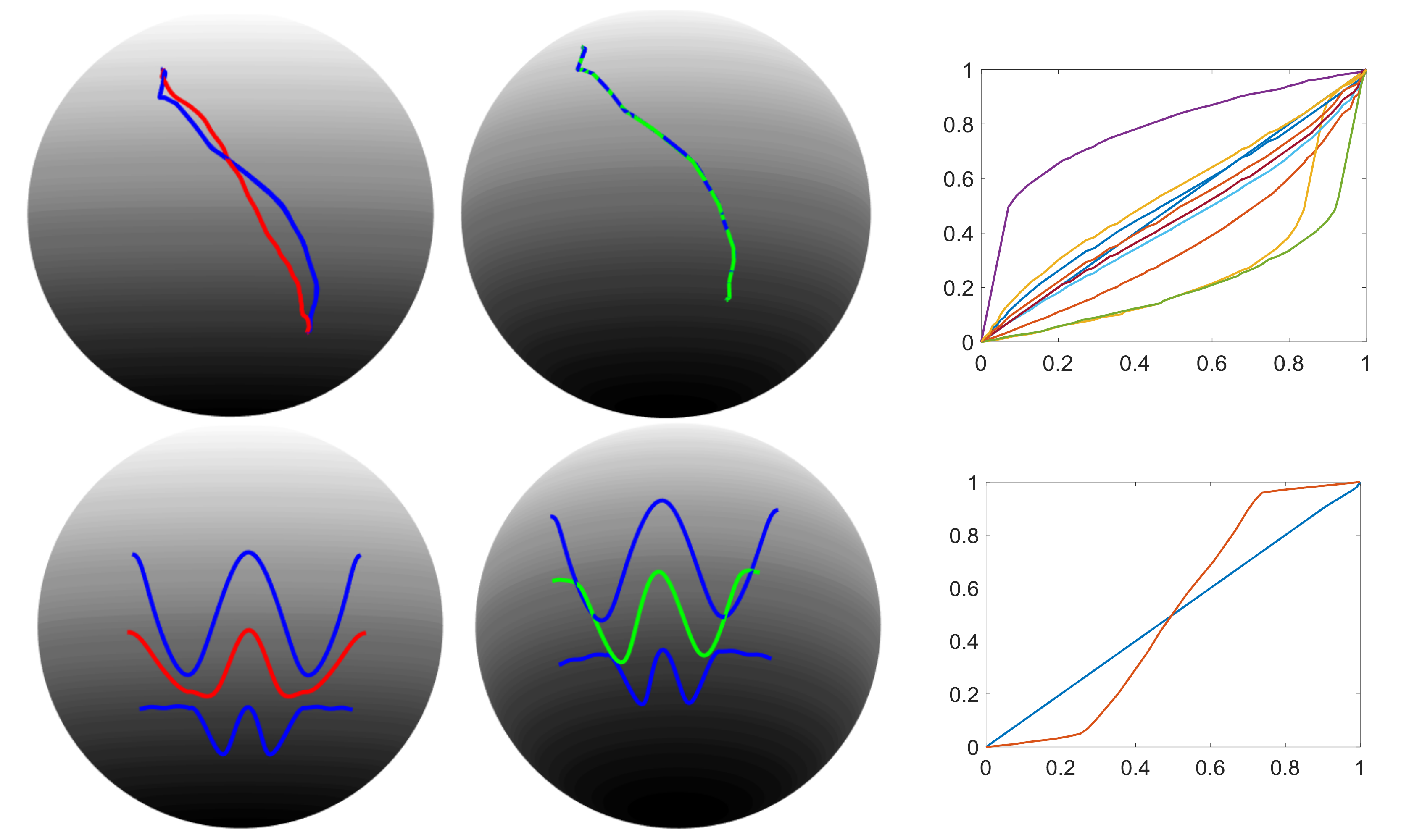}\\
\end{tabular}
\caption{Comparison of Euclidean mean and Karcher mean. The two rows denote two 
different examples. }\label{Karchermean} 
\end{center}
\end{figure}

In the next experiment, we start from a trajectory, say  $\alpha_1$, and then introduce arbitrary phases
 to obtain $ \alpha_1 \circ \gamma$. 
Time warping a trajectory  does not change its amplitude, but changes its phase. 
Fig. \ref{Karchermean} shows ten such trajectories in the upper left panel,  
drawn in blue lines, and their Euclidean mean (cross-sectional mean) in the red line. 
One can see that the Euclidean mean has different amplitude 
from the original trajectories despite the original ones having the same amplitude. 
Now, if we perform phase-amplitude separation, 
and compute the amplitude mean under $d_a$, the 
result is shown in the  middle upper panel (in green color). The mean trajectory has the exact
same amplitude as $\alpha_1$, and
the relative phase ${\gamma_i^*}$ is shown in the last column.

Next, we simulate two spherical trajectories 
with two ``bumps" each, as shown in the bottom-left panel (in blue lines) of 
Fig. \ref{Karchermean}. and the red line is their cross-sectional mean. 
Then we compute their amplitude mean which is shown in green line in the bottom-middle panel. 
One can see that the mean amplitude (after phase-amplitude separation) 
better preserves the bump features, as compared to the Euclidean mean.  
The relative phase  $\{\gamma_i^*\}$ is shown in the bottom-right panel. 
 
\subsection{Real Data}
In this section, we illustrate our framework on two  real datasets: bird migration data \citep{Kochert2011} and hurricane tracks \citep{HURDAT2}. 

\begin{enumerate}
\item {\bf Bird Migration Data:} The bird migration data in  \citet{Kochert2011}  contains $35$ migration trajectories of Swainson's Hawk, observed during $1995$ to $1997$. The migration of each 
Swainson's Hawk was tracked using satellite tag and their locations  was recorded by the satellite every $1$-$4$ days. 
\item {\bf Hurricane Tracks:}  We use the
Atlantic hurricane database (HURDAT2) \citep{HURDAT2} to get hurricane tracks. HURDAT2  is a tropical cyclone historical database containing hurricanes starting from north Atlantic ocean and Gulf of Mexico from 1851 to 2015. The database contains six-hourly information on the locations, maximum winds, central pressures and so on for each of the relevant hurricane.
\end{enumerate}

\begin{figure}
\begin{center}
\begin{tabular}{|c|}
\hline
{Hurricane example}\\
\includegraphics[height=1.4in]{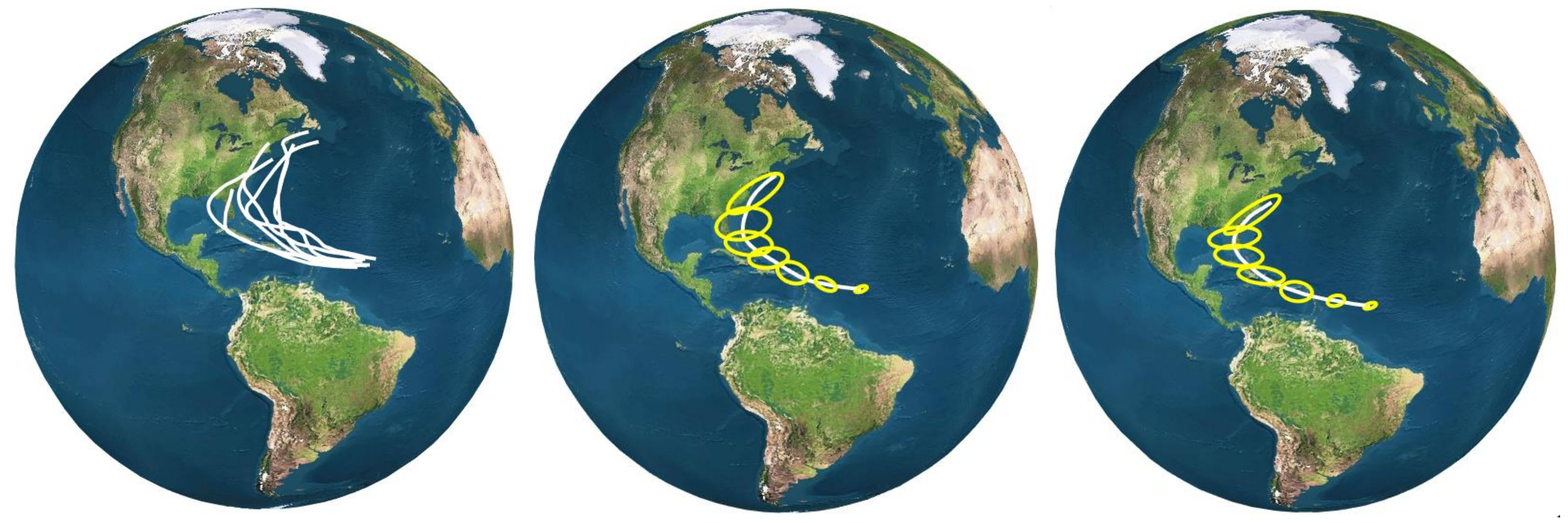}\\
\hline
{Bird migration example}\\
\includegraphics[height=1.4in]{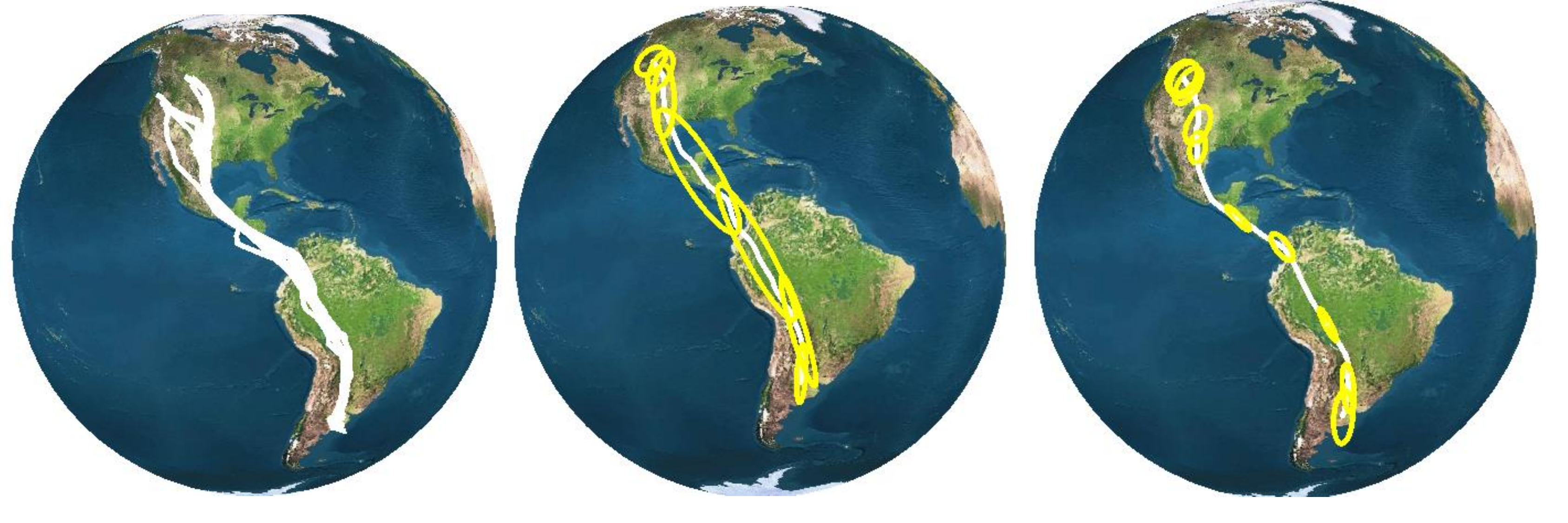}\\
\hline
\end{tabular}
\caption{Examples of Karcher mean trajectory in real data. The first column shows the original data, the second column shows the cross-sectional mean and the third column shows the mean after temporal alignment.  The yellow ellipsoids shows the cross-sectional variance for data along the mean trajectory.  }\label{Karchermean_real} 
\end{center}
\end{figure}

\begin{figure}
\begin{center}
\begin{tabular}{cc}
\includegraphics[height=1.2in]{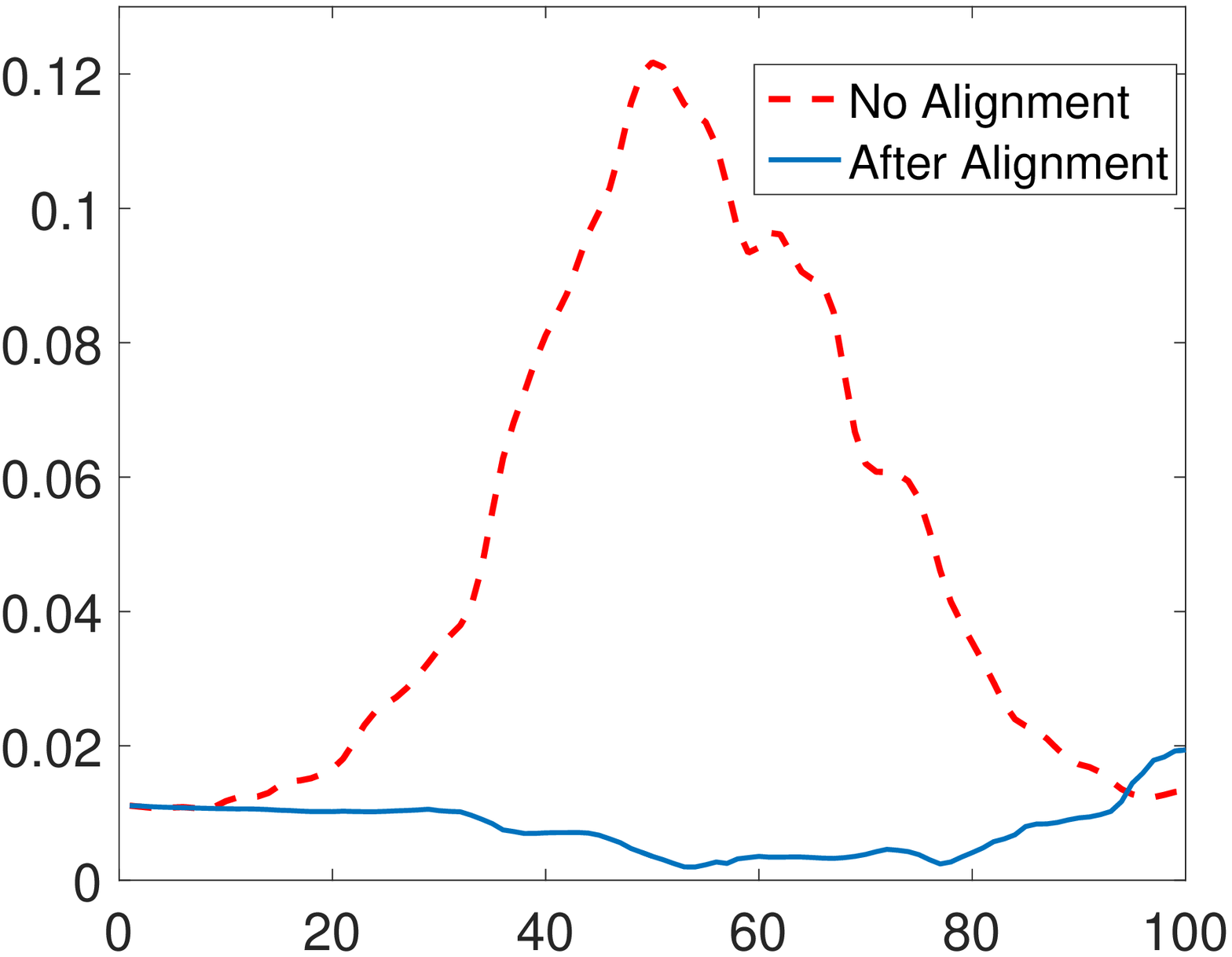}&
\includegraphics[height=1.2in]{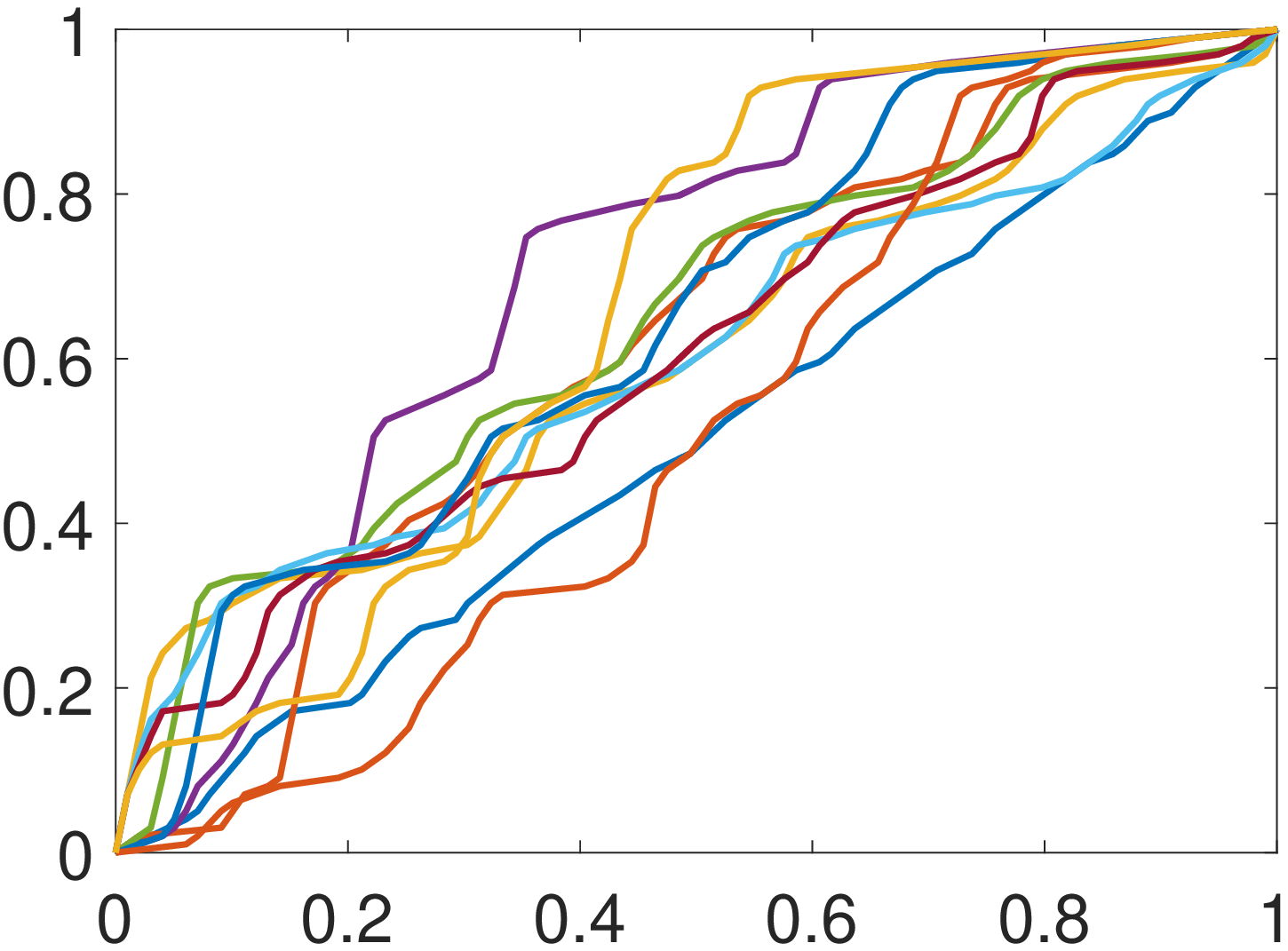}\\
\includegraphics[height=1.2in]{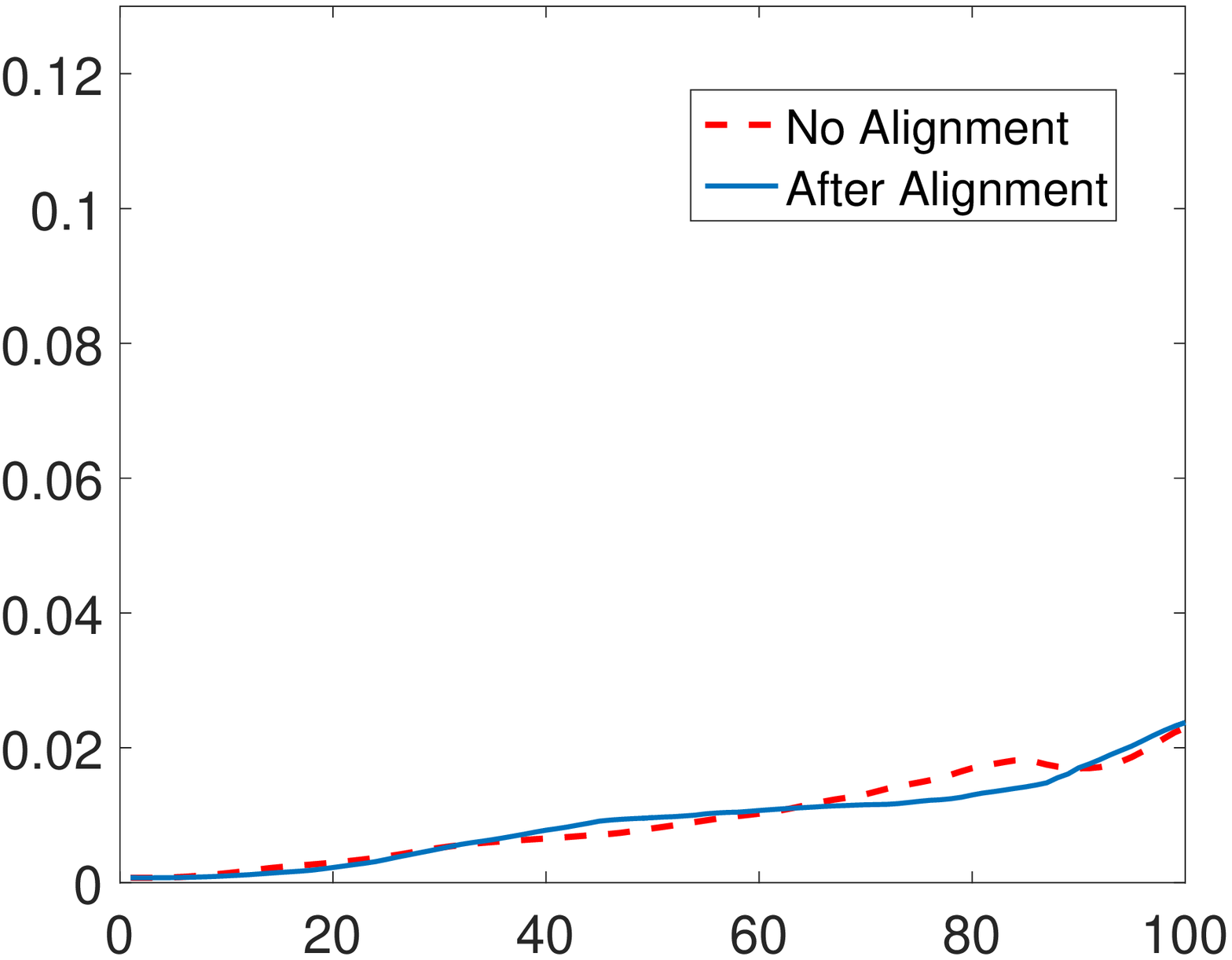}&
\includegraphics[height=1.2in]{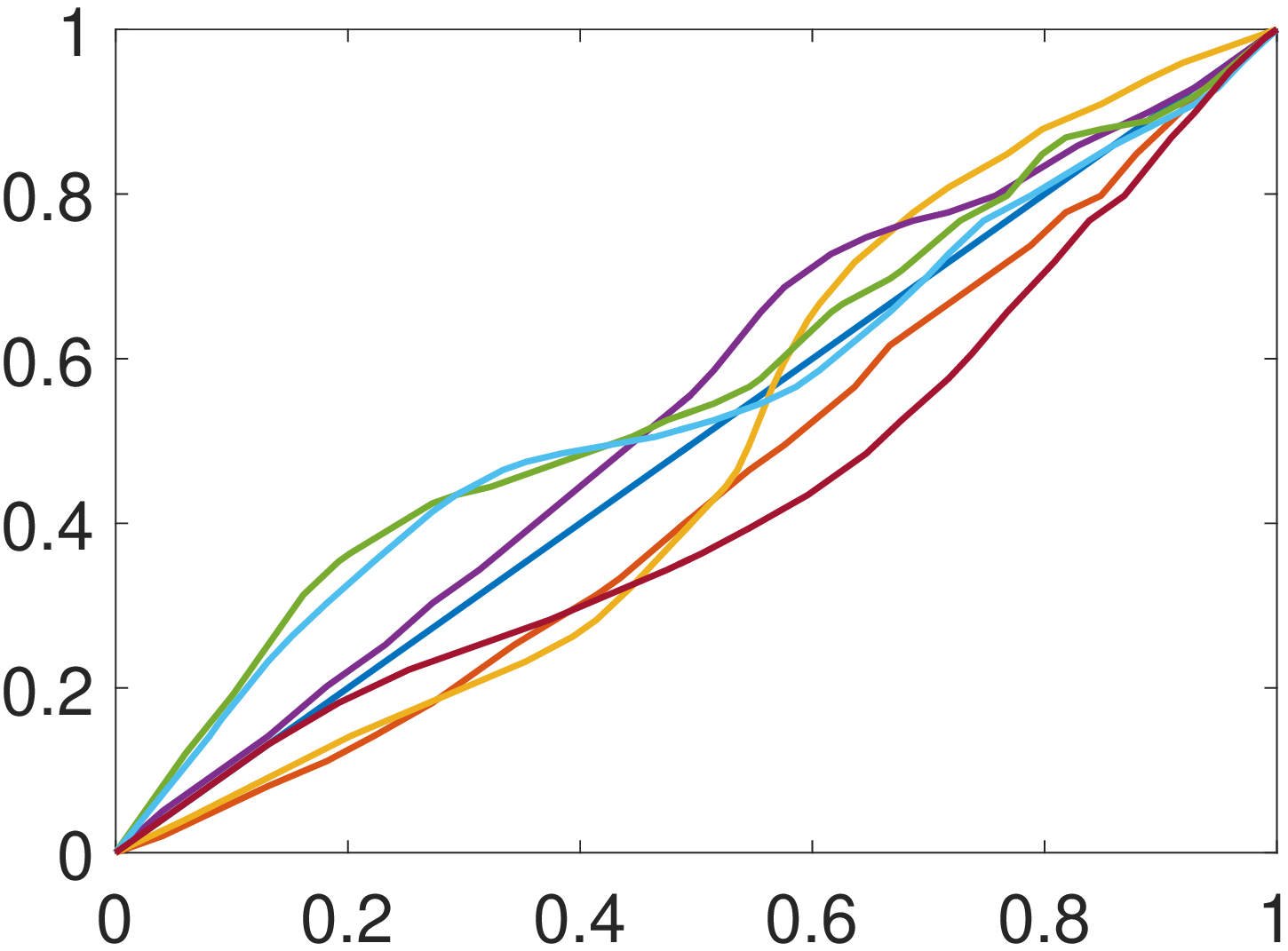}\\
\end{tabular}
\caption{Variance of the amplitude components and phase functions for bird migration (first row) and hurricane (second row) examples. } \label{fig:karmean} 
\end{center}
\end{figure}

We randomly choose $7$ hurricane trajectories and $10$ migration trajectories, and calculate their Euclidean means and 
mean amplitudes (Karhcer mean with phase removal). The results are shown in Fig. \ref{Karchermean_real}. The first column shows the original data, second column shows their Euclidean means and the third column shows the Karcher means of the amplitude components. Separating the phase components reduces temporal variance inside the trajectories and makes the remaining amplitude components compact. 
To emphasize this point, we calculate the 
cross-sectional variance at some discrete sampling points along the mean trajectory, 
say $\{t_1,...,t_7\}$. At each $t_i$, the cross-sectional variance is a $3\times3$ matrix, and we use its first two principal directions to
display this variance matrix as a tangential ellipsoid. Fig. \ref{Karchermean_real} (columns two and three) show these 
ellipsoids along the mean trajectories before and after phase separation.  
Another way to illustrate variance reduction due to phase removal is shown in Fig. \ref{fig:karmean} column one, where the x-axis is time 
and y-axis is the trace of the variance matrix. The relative phase components are shown in the second column in Fig.\ref{fig:karmean}.  From Fig. \ref{Karchermean_real} and Fig. \ref{fig:karmean} one can see that before alignment, the variance at each sampling point is mainly along the trajectories, especially in the bird migration case, which means that the birds are flying at different speeds, 
and this inflates variance tremendously. The alignment process separates the phase from amplitude, 
and retains only the amplitude differences. Also, we note that 
phase variability in the  hurricane data is relatively small.   

\begin{figure}
\begin{center}
\begin{tabular}{|c|c|}
\hline
{First two PCs for $u$ with $w = {\bf 0}$ ($100\%$) } & First two PCs for $w$ with $u = {\bf 0}$ ($41.89\%$) \\
\includegraphics[height=1.50in]{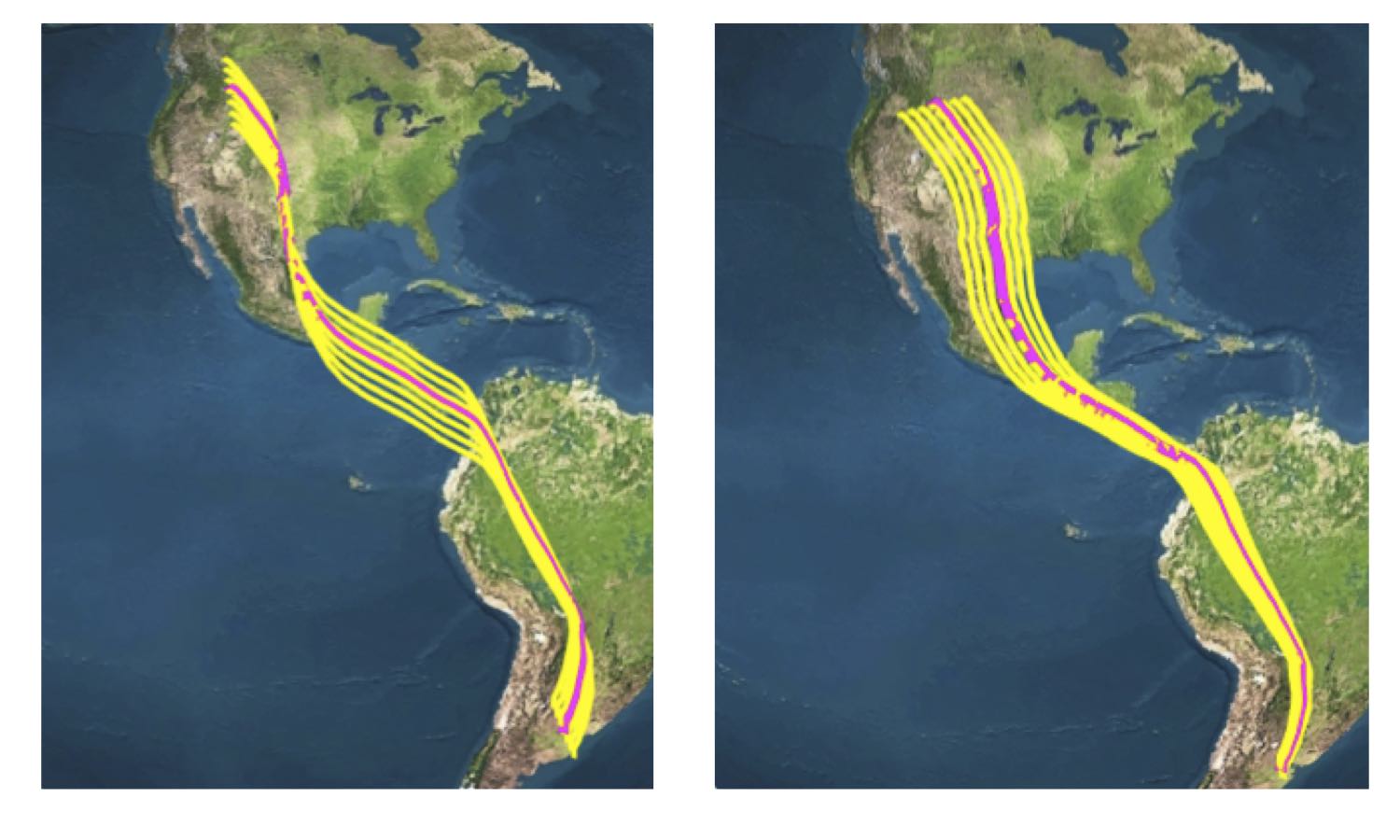} &
\includegraphics[height=1.50in]{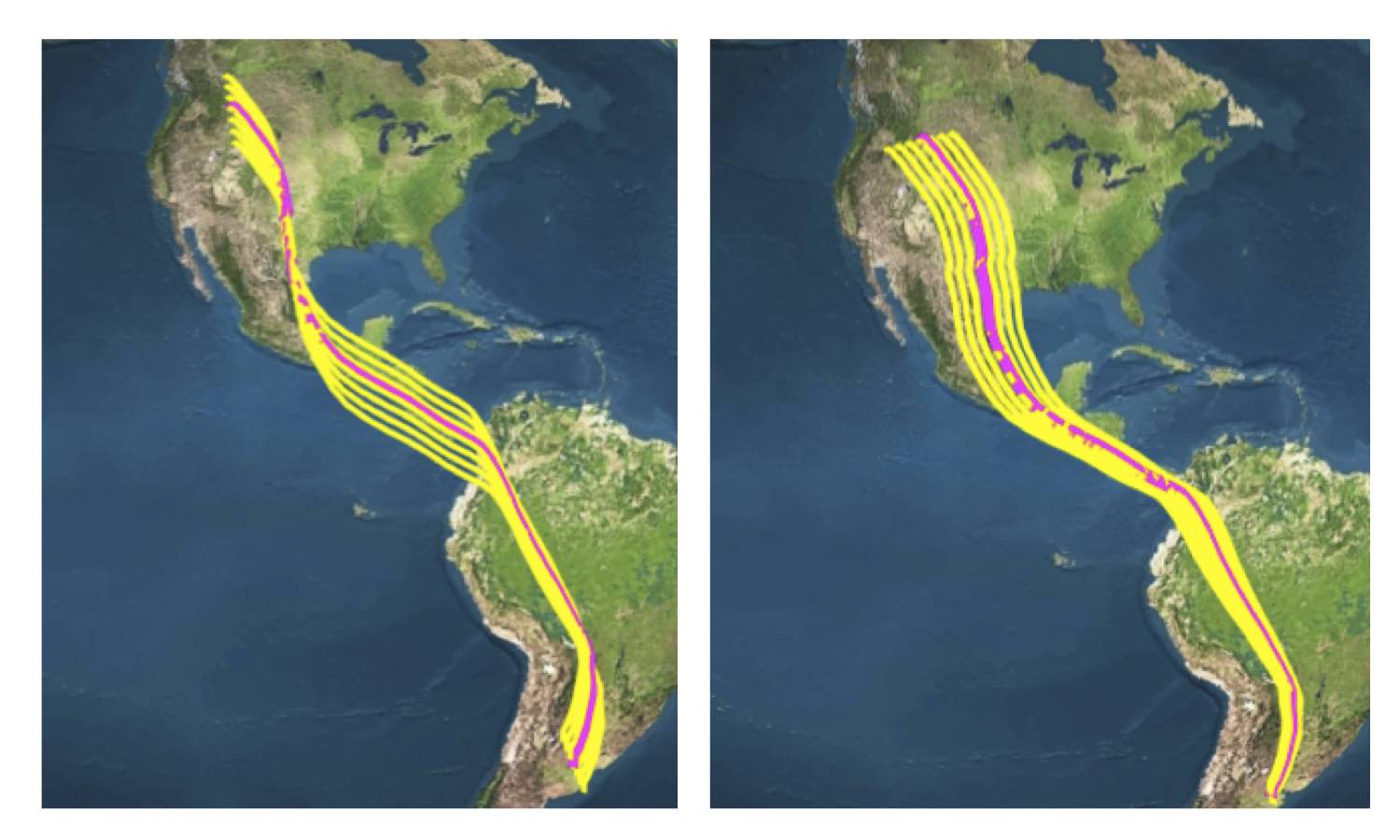}\\
\hline
First two PCs for $u$ with $w = {\bf 0}$  ($100\%$) & First two PCs for $w$ with $u = {\bf 0}$ ($69.44\%$) \\
\includegraphics[height=1.50in]{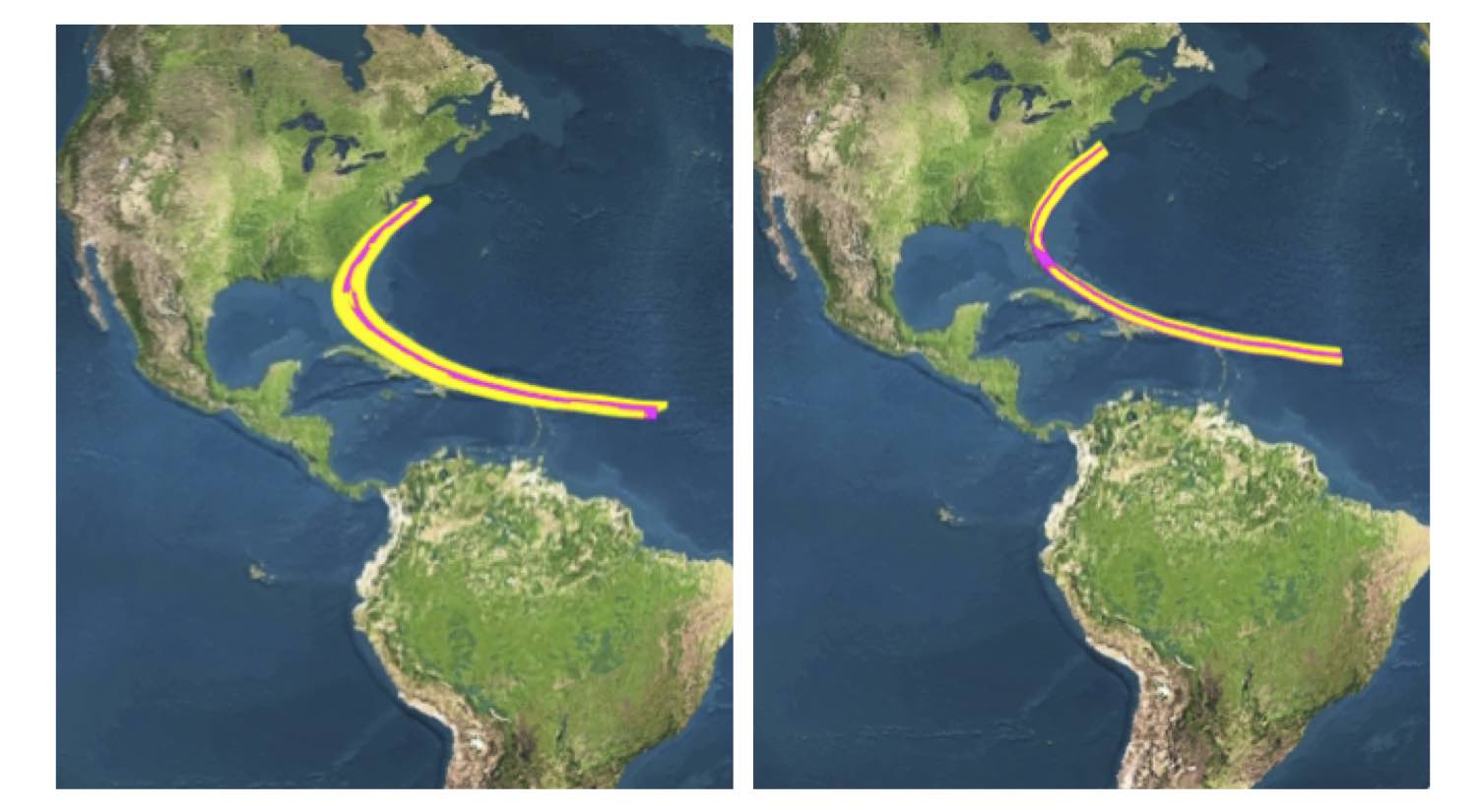}&
\includegraphics[height=1.50in]{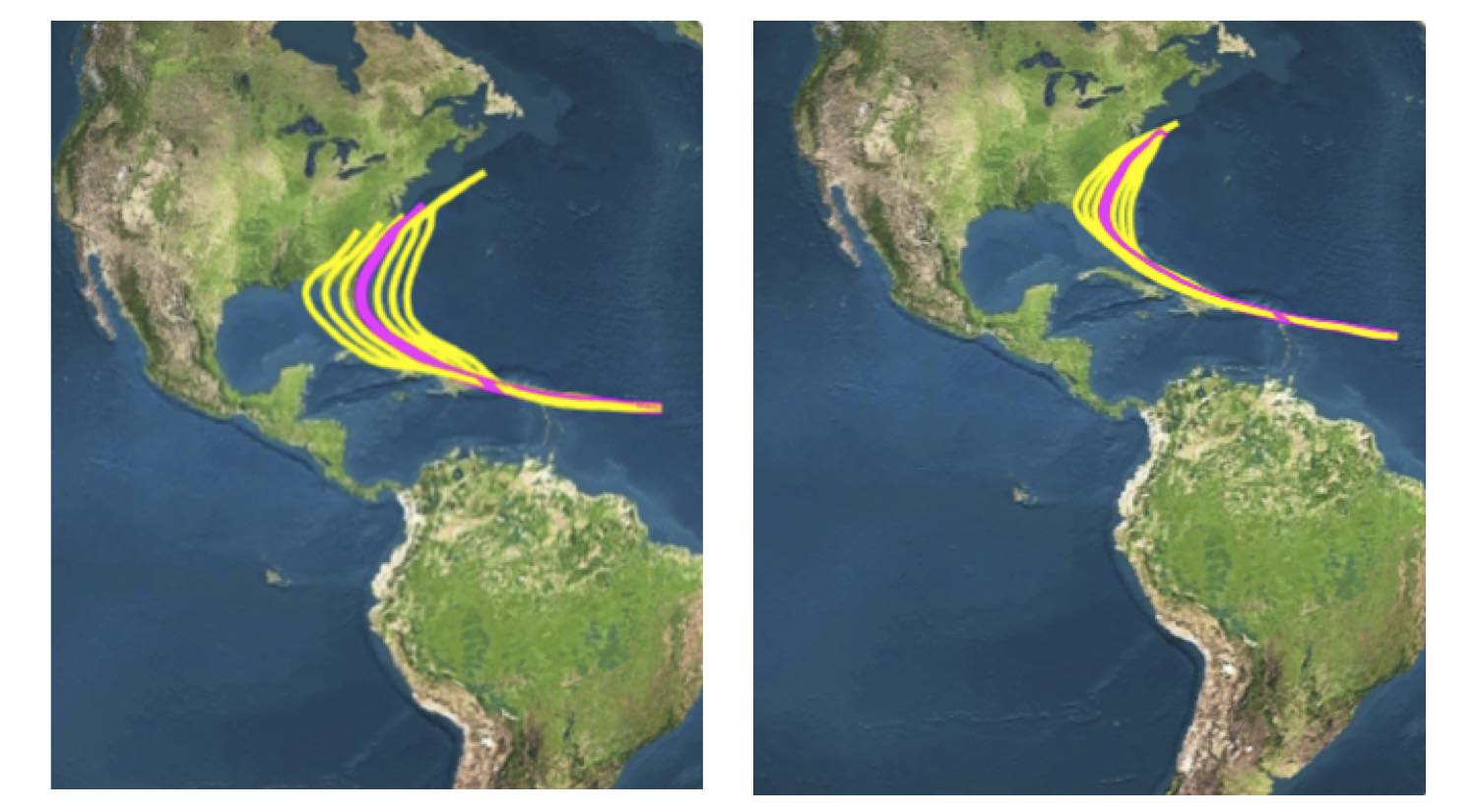}\\
\hline
\end{tabular}
\caption{PCA for bird migration and hurricane data in Fig. \ref{Karchermean_real}. The first row shows result for bird migration data and the second row shows result for hurricane data. }\label{fig:pcabird} 
\end{center}
\end{figure}

We also use the method described in Section \ref{sec:pca} to perform 
mfPCA on amplitudes in the two datasets. 
We show the first two principal directions for the two datasets in Fig. \ref{fig:pcabird}. In each row, in the left panel, we let the second component $w = {\bf 0}$ and show the two modes of variation in the first component $u_i$, and in the right panel, we let $u={\bf 0}$ and show the first two modes of variation in $w_i$. The middle curve in magenta color, with $\tau = 0$, is the mean trajectory.  In the parentheses in each row, we show the percentage of variation that was explained by the first two PCs.

\begin{figure}
\begin{center}
\includegraphics[height=2.0in]{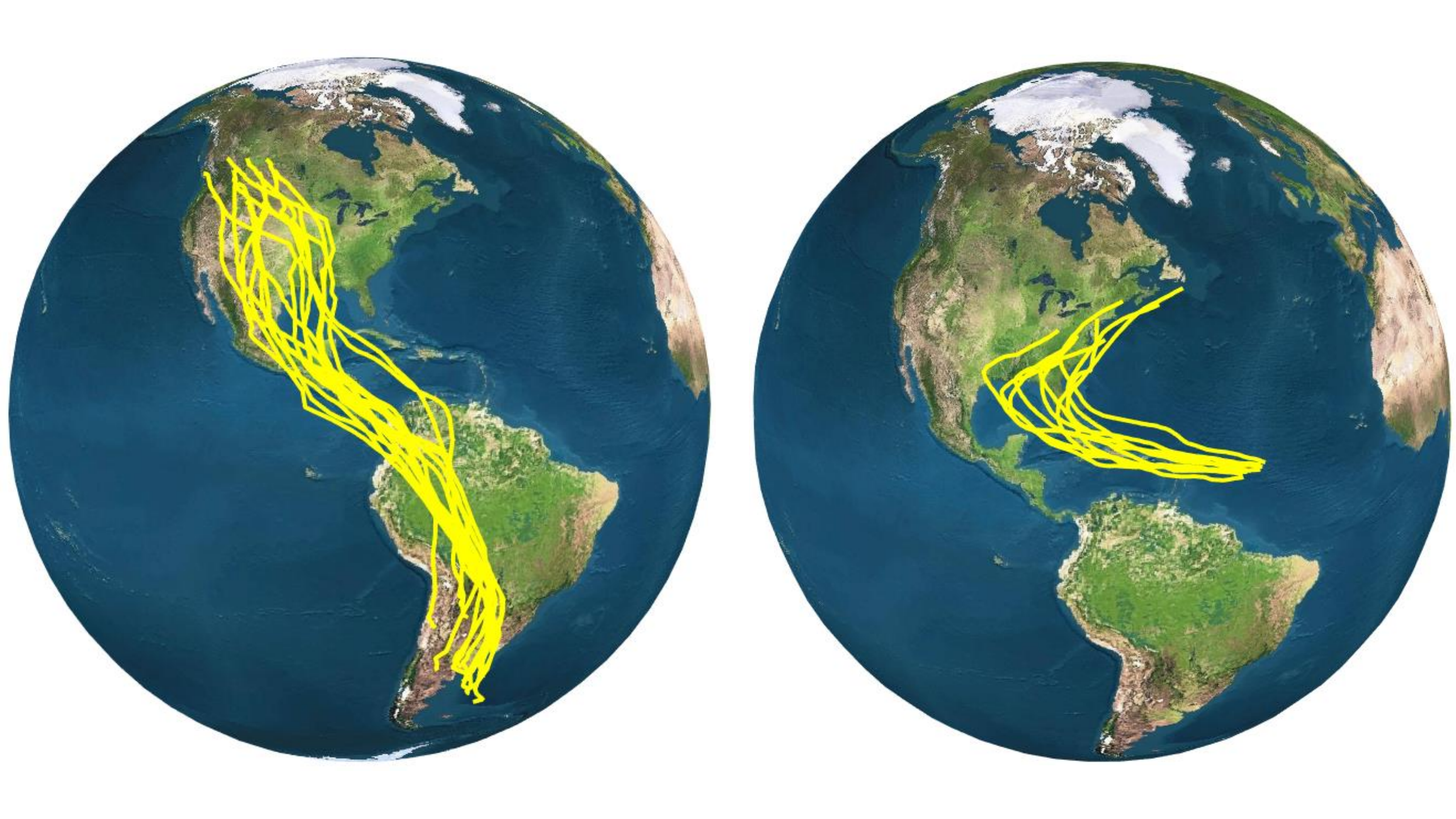}
\caption{Examples of random sampling from the wrapped Gaussian model. The left panel shows samples from bird migration data and the right panel shows the samples from the hurricane trajectories. }\label{fig:randomsample} 
\end{center}
\end{figure}

To capture the distributions of the bird migration and hurricane subsets, we use the wrapped Gaussian model described in Section \ref{sec:Gmodel} to generate random samples. Fig. \ref{fig:randomsample} displays some random samples from the wrapped Gaussian distribution on $\cC$.

\subsection{Clustering of Hurricane Trajectories}

\begin{figure}
\begin{center}
\begin{tabular}{cc}
\includegraphics[height=2in]{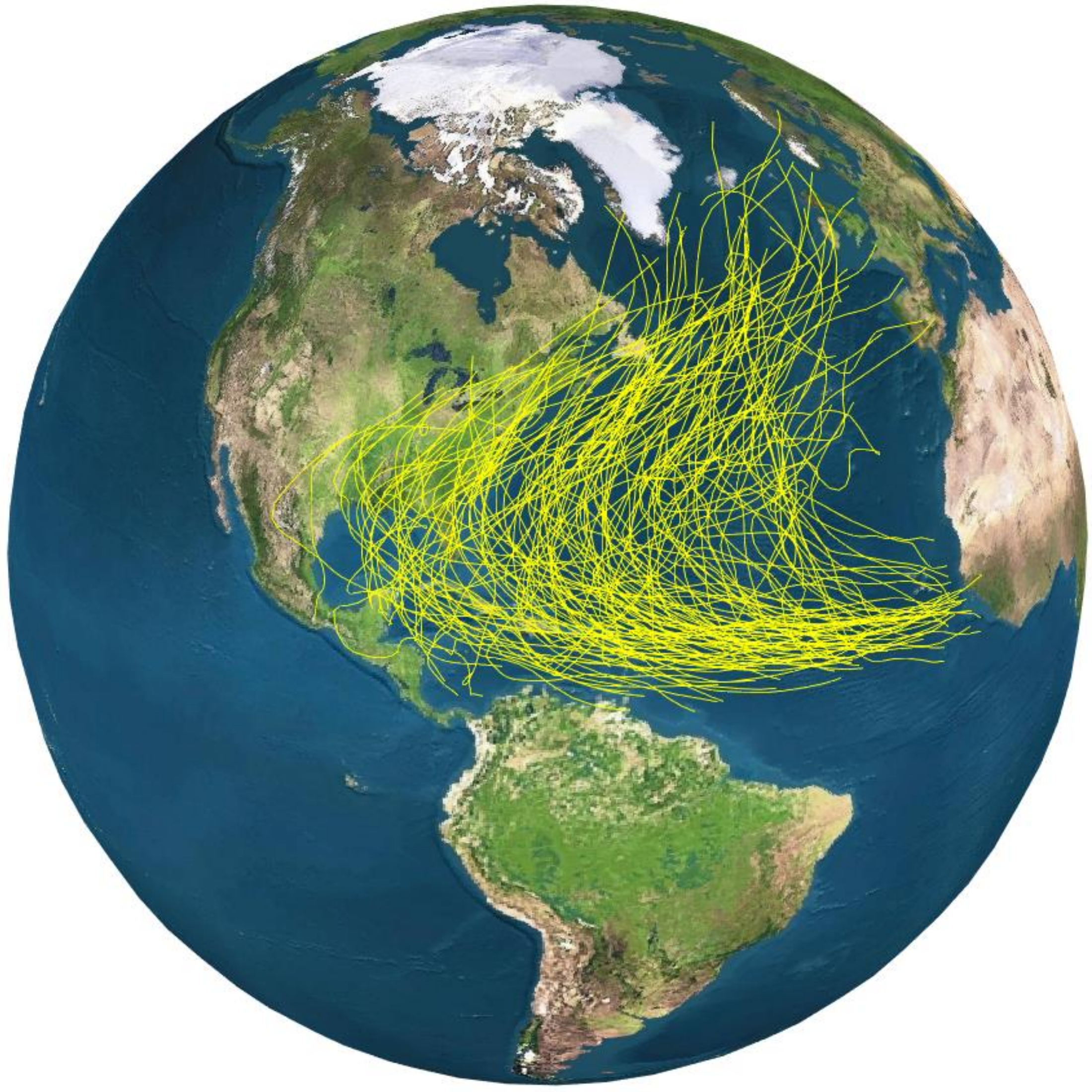} &
\includegraphics[height=2in]{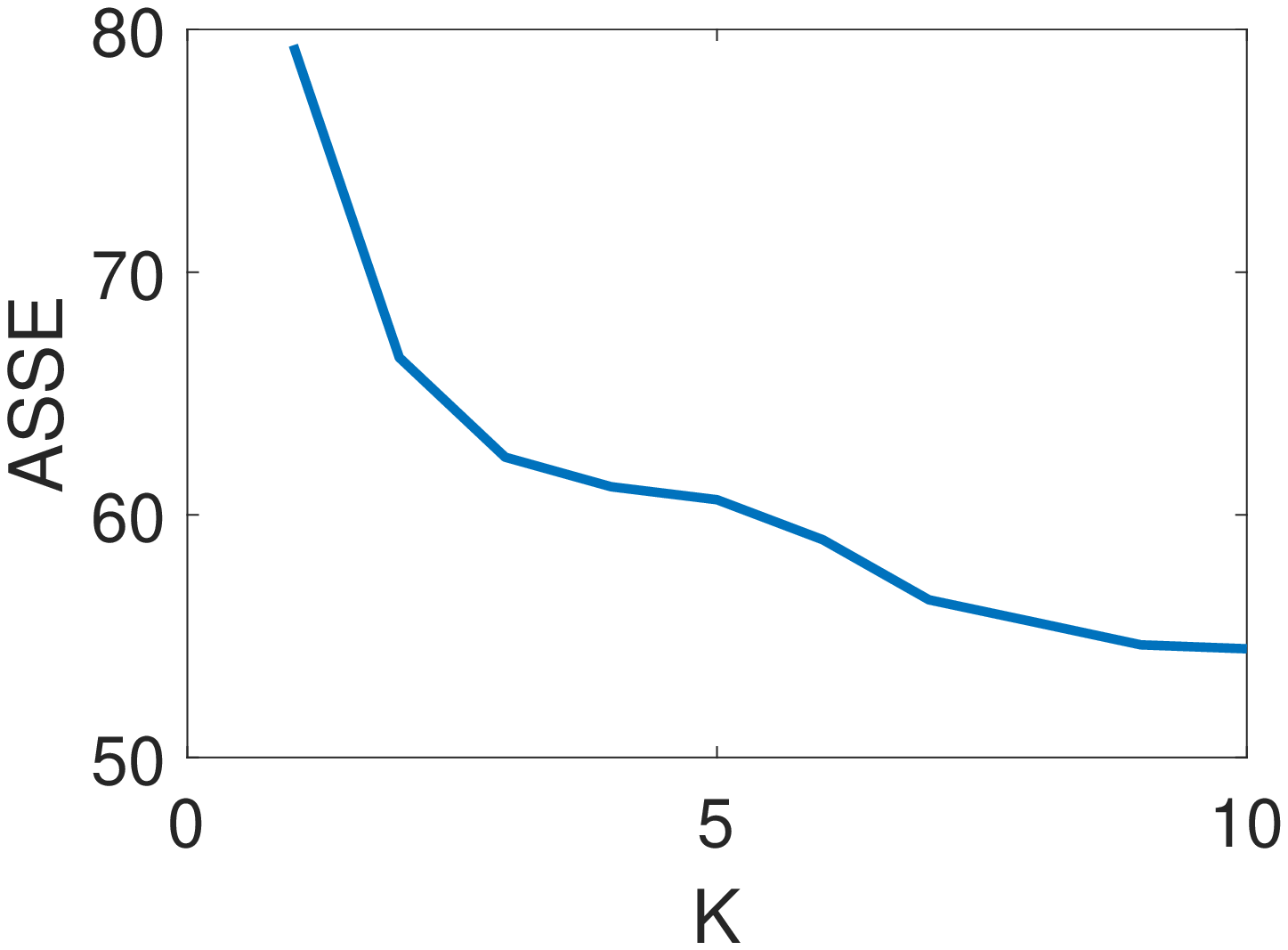} \\
(a) & (b) \\
\end{tabular}
\caption{ (a) 138 selected hurricanes starting before latitudes of of $20^o$N and ending after $35^o$N  (b) The averaged sum of squared error versus the chosen $k$ for $k$-means method.}\label{fig:138hurricanes} 
\end{center}
\end{figure}

Next we consider the problem of clustering of hurricane trajectories, 
in a manner that is invariant to their phase variability. 
For this experiment, we extract all those
 trajectories that start before latitude of $20^o$N and end after $35^o$N from  the database of 
 trajectories recorded during 1969-2014, similar to the data used in \citet{kendall-arxiv:2014}.  This extraction 
 results in $138$ trajectories and Fig. \ref{fig:138hurricanes} (a) shows some examples. 

To cluster these tracks, one of the simplest methods is $k$-means clustering algorithm, using 
the amplitude distance $d_a$ and use Algorithm \ref{algo5} to calculate a mean trajectory under $d_a$. 
We use Lloyd's algorithm \citep{Lloyd:2006} for $k$-means: beginning with a random initial set of $k$ trajectories
serving as cluster centroid trajectories, the algorithm first associates each trajectory to the closest cluster centroid trajectory (measured
by Eqn.\ref{eqn:two_para}), and then replacing each cluster centroid trajectory by the
computed Karcher mean trajectory (calculated by Eqn. (\ref{eqn:karchermean})) for the cluster. The algorithm is
iterative and is guaranteed to coverage locally.  However, $k$-means algorithm requires us to provide the number of clusters $k$, which in unknown for the hurricane trajectories. Although some methods, such as $G$-means \citep{NIPS2003_2526}, $X$-means \citep{Pelleg2000}, provide some algorithms to decide $k$, but they only work in the Euclidean cases. Here we use the classical ``Elbow'' method to decide $k$. The averaged sum of squared error (ASSE) for each value of $k$ is calculated and plotted. ASSE decreases as $k$ gets larger,  and the Elbow method is to choose the $k$ at which the ASSE stops decreasing abruptly. In Fig. \ref{fig:138hurricanes} (b), we show the plot the ASSE versus k, and we choose $k=3$. 

\begin{figure}
\begin{center}
\includegraphics[height=3.7in]{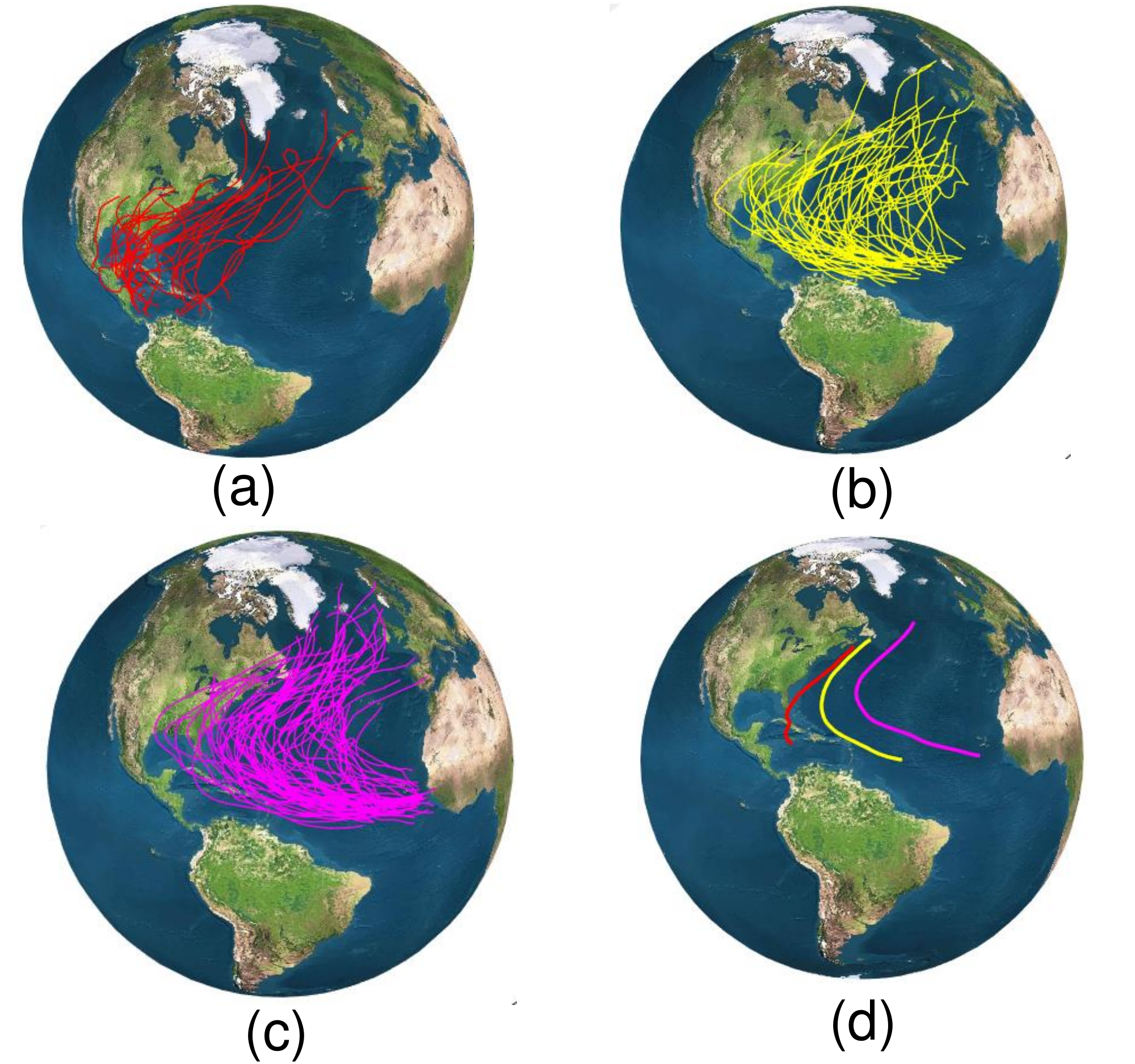}
\caption{ K-means clustering result. (d) shows the mean tracks of the three clusters in (a), (b) and (c). }\label{fig:kmeanclusteringr} 
\end{center}
\end{figure}

With a fixed $k$, we apply the $k$-means algorithm to cluster $138$ hurricanes. However, it is well known that $k$-means is not robust to different initializations and results in a local minimum. As a result, $k$-means method might have different results with different initializations. To tackle this problem, we propose a vote-based $k$-means method, i.e. we run $k$-means algorithm multiple times, 
with different initial conditions, and the final result is based on the average of these $k$-means results. 
Similar to \citep{Zhang2015171}, for each $k$-means clustering result, we use a binary matrix $B$ to represent the clustering configuration such that $B(i,j) = 1$ if $i$-th and $j$-th elements are from the same cluster. Given $n$ different $k$-means clustering results, denoted as $B_i$ for $i=1,...,n$, the final $B$ is obtained by calculating the extrinsic mean of $B_i$s using Algorithm 2 in \citet{Zhang2015171}. In Fig. \ref{fig:kmeanclusteringr}, we show the final clustering result based on the voting of $50$ $k$-means clustering results (by setting $k=3$). 

\begin{figure}
\begin{center}
\includegraphics[height=1.6in]{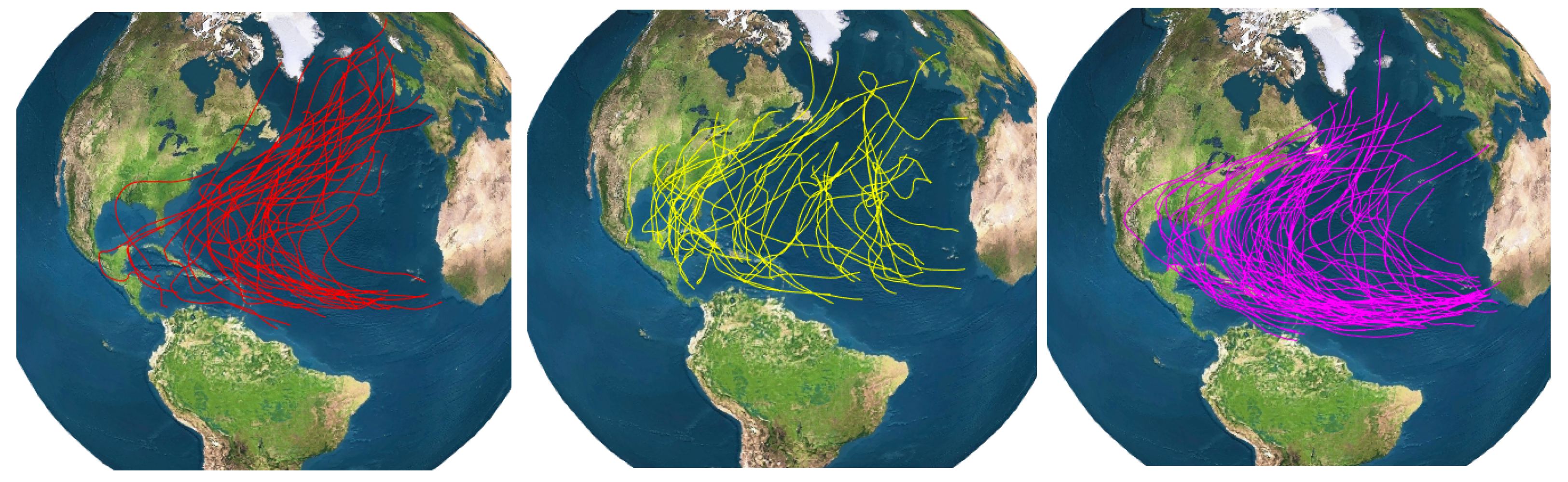}
\caption{Clustering result by treating hurricane trajectories as curves in $\real^3$.    }
\label{fig:3dcurvesclustering} 
\end{center}
\end{figure}

To validate our $k$-means clustering result, we employe the stochastic simulated annealing clustering algorithm in \citet{Srivastava2005} to perform the clustering. Pairwise amplitude distances $d_a$ between 
$138$ hurricane tracks are calculated, and then an annealing method is used to re-arrange the tracks into $k=3$ clusters.  
This clustering of tracks is found to agree with 
the vote-based $k$-means result for $96\%$ of the tracks, thus validating our clustering results. 
As another  comparison, we treat the hurricane trajectories as regular curves in $\real^3$, and then we can use the elastic shape analysis framework in \citep{anuj2011} to calculate the pairwise distance between the shapes of trajectories (by removing the translation, rotation, scaling and re-parameteriation). Using this distance, we can perform the clustering. Fig. \ref{fig:3dcurvesclustering} shows the clustering result using $k$-means method by setting $k=3$. Comparing with the clustering result of the proposed framework (in Fig. \ref{fig:kmeanclusteringr}) we can see that the proposed method has more meaningful result: the hurricanes starting from Gulf of Mexico and Caribbean Sea tend to move along the east coast and are less curved comparing with hurricanes starting in the North Atlantic Ocean; the hurricanes starting from the part of North Atlantic Ocean near Africa tend to have long lengths and are left curved; hurricanes starting from the part of North Atlantic Ocean near South America have shapes in between of the pervious two classes.   

\section{Conclusion}
In summary, we have proposed a principled approach for phase-amplitude separation of spherical trajectories, 
using a metric that has appropriate invariance properties. 
Each spherical trajectory is represented by a pair: a starting point and a curve on the tangent space of the starting point, called {\it TSVRC}. 
Such representation forms a vector bundle and allows separating of phase from the amplitude of the trajectory. Using simple 
geometry of $\s^2$, we have defined fast algorithms to calculate geodesics between elements of this vector bundle. 
Explicit expressions for exponential map and inverse exponential map are also developed to facilitate the analysis of multiple trajectories: calculating the Karcher mean, separating phase-amplitude for multiple trajectories and performing PCA on the aligned trajectories.  Both simulated and real data are used to validate the developed procedures and demonstrate the advantages of analyzing and modeling the trajectories after alignment.
\appendix

\makeatletter   
 \renewcommand{\@seccntformat}[1]{APPENDIX~{\csname the#1\endcsname}.\hspace*{1em}}
 \makeatother

\section{Riemannian Structure on $\s^2$}
To perform the trajectory analysis on the manifold $\s^2$, one needs a Riemannian structure on this manifold. Specially, we need the following tools: (1) {\it geodesic} between two points on the manifold, (2) {\it parallel transport} of tangent vectors along the geodesic path, (3) {\it exponential map}, (4) {\it inverse exponential map} and (5) {\it Riemannian curvature tensor}. 

We use a simple Euclidean inner product as the Riemannian metric on $\s^2$: for any $v_1,v_2 \in T_p(M)$, the metric is defined to be: $\left<v_1,v_2\right> = v_1^t v_2$. For any two points $p,q \in \s^2 (p \neq q)$ and a tangent vector $v \in T_p(\s^2)$, we have the following closed solutions for the tools we need: 

\begin{enumerate}
\item {\it Geodesic:} The geodesic between $p$ and $q$ is the great circle connecting them: $\alpha(t) = {1}/ {\sin(\theta)} {(\sin(\theta(1-t))p + \sin(\theta t)q)}$,  where $\theta$ is determined by $\cos(\theta) = \left<p,q\right>$ and $0<\theta<\pi$

\item {\it Parallel Transport:} The parallel transport $(v)_{p \to q}$ along the shortest geodesic (i.e. great circle) from $p$ to $q$ is given by $v - 2\left<v,q\right>(p+q)/|p+q|^2.$

\item {\it Exponential Map:} The exponential map $\exp_p(v)$ is $\cos(\|v\|)p + \sin(\|v\|)v/\|v\|$.
\item {\it Inverse Exponential Map:} The inverse exponential map $\exp_{p}^{-1}(q)$ is $(q-p\cos(\theta))\theta/\sin(\theta)$, $\theta = \cos^{-1}(\left<g_1,g_2 \right>)$. 

\item {\it Riemannian Curvature Tensor:} For three tangent vectors $x,y,z$ on $T_p(\s^2)$, the Riemannian curvature tensor $R(x,y)(z) = \left<y,z\right>x - \left<x,z\right>y = -(x\times y \times z)$, where $\left<\cdot\right>$ denotes the ordinary inner product, and $\times$ denotes the cross product.

\end{enumerate}

\section{Two properties for geodesics on $\cC$} \label{app:twopt}
Suppose that $(\beta(t),q(t))$ is a geodesic on $\cC$, where $t\in I=[0,1]$. Let $L$ denote the length of the path $\beta:I\to \s^2$. Let $\tilde\beta:I\to \s^2$ be a constant speed re-parametrization of $\beta$. For each $t_0\in I$, define 
$Z_{t_0}:\ltwo(I,T_{\tilde\beta(0)}(\s^2))\to \ltwo(I,T_{\tilde\beta(t_0)}(\s^2))$ by parallel transporting each tangent vector along $\tilde\beta(t)$. Then we can define

$$Z:[0,L]\times \ltwo(I,T_{\tilde\beta(0)}(\s^2))\to \cC$$

by $Z(s,q)=(\tilde\beta(s/L),Z_{(s/L)}(q))$. A routine verification shows that if we put the standard product Riemannian metric on $[0,L]\times \ltwo(I,T_{\tilde\beta(0)} (\s^2))$, then $Z$ is an isometric immersion. Since our original geodesic $(\beta(t),q(t))$ is contained in the image of $Z$, its inverse image under $Z$ in  $[0,L]\times L^2(I,T_{\tilde\beta(0)} (\s^2))$ must be a geodesic. But since this latter space is Euclidean (i.e., we are using the same Riemannian metric at each point), it follows that the inverse image of our geodesic must be a straight line in this space. It follows immediately that $\beta(t)$ must have constant speed and $q(t)$ must be covariantly linear.

\section{Proof of Lemma 2}\label{app1}

Let $\beta$ be the shortest geodesic joining $p_1$ to $p_2$ and let $L:T_{p_1}(\s^2)\to T_{p_2} (\s^2)$ be the parallel translation map induced by $\beta$. Let $\zeta$ be any other path from $p_1$ to $p_2$ that is disjoint from $\beta$. The Gauss Bonnet Theorem states that the angle of rotation of the parallel translation map $T_{p_2} (\s^2)\to T_{p_2} (\s^2)$ induced by the concatenation $\zeta^{-1}*\beta$ is equal to the integral of the Gaussian curvature over the region enclosed by the loop $\beta \cup \zeta$. Since the Gaussian curvature of $\s^2$ equals +1 at every point, this implies that this angle of rotation is equal to the area enclosed by the loop. However, it is well known that of all curves that enclose a given area, a circle is the shortest! From this, it is easy to prove that if part of your loop is already given (by the geodesic, as in this case), then the shortest way to fill in the rest of your arc to enclose a given area is by a circular arc. This proves Lemma 2.

\section{Proof of Theorem 1}\label{app2}

Let $\gamma\in\Gamma$ be a warping function, and let $\gamma$ act on the space $\cC$ by 
$(x,q)*\gamma=(x,(q*\gamma))$. The differential of this action is the map 
$T_{(x,q)}(\cC)\to T_{(x,(q*\gamma))}(\cC)$  given by $(u,w)\mapsto (u,w*\gamma)$. We prove that this differential preserves our Riemannian inner product (Eqn. \ref{eqn:inner}) as follows: let   $(u_1,w_1)$ and $(u_2,w_2)$ be two tangent vectors on $T_{(x,q)}(\cC)$; it follows that 
\begin{eqnarray*}
\left< (u_1,w_1*\gamma),(u_2,w_2*\gamma)\right> & = & u_1\cdot u_2 + \int_0^1 w_1(\gamma(t)) \sqrt{\dot{\gamma}(t)} w_2(\gamma(t)) \sqrt{\dot{\gamma}(t)} dt \\
 & = & u_1\cdot u_2 + \int_0^1 w_1(\gamma) w_2(\gamma) d\gamma \\
 & = & u_1 \cdot u_2 +  \int_0^1 w_1(s) w_2(s) ds \\
 & = & \left< (u_1,w_1),(u_2,w_2)\right>
\end{eqnarray*}
Since $\Gamma$ acts on $\cC$ by isometries, i.e. preserving the Riemannian inner product, it follows immediately that it takes geodesics to geodesics, and preserves geodesic distance. It also follows that it preserves the baselines of these geodesics, i.e. $\beta^*$. 

\bibliography{bibfile}
\bibliographystyle{rss}

\end{document}